\documentclass[letterpaper, 10 pt, conference]{ieeeconf}  % Comment this line out if you need a4paper

\IEEEoverridecommandlockouts                              % This command is only needed if 
\usepackage{amsmath} % assumes amsmath package installed
\usepackage{amssymb}  % assumes amsmath package installed
\usepackage{amsmath}
\usepackage{algorithm}
\usepackage{algorithmicx}%, algorithmic}
\usepackage{algpseudocode}
\usepackage{xcolor}
\usepackage{colortbl}
\usepackage{amssymb}
\usepackage{graphicx}
\usepackage{newfloat}
\usepackage{listings}
\usepackage{booktabs}
\usepackage{multirow}
\usepackage{makecell}
\usepackage{subcaption}
\usepackage{cleveref}

\title{\LARGE \bf
Deploying Ten Thousand Robots: Scalable Imitation Learning for Lifelong Multi-Agent Path Finding
}

\author{
He Jiang$^{1*}$, Yutong Wang$^{2*}$, Rishi Veerapaneni$^{1}$, Tanishq Duhan$^{2}$, Guillaume Sartoretti$^{2}$, Jiaoyang Li$^{1}$ % <-this % stops a space
\thanks{*The first two authors contribute equally.}% <-this % stops a space
\thanks{$^{1}$These authors are with the Robotics Institute, Carnegie Mellon University, USA. \{hejiangrivers,vrishi,jiaoyangli\}@cmu.edu.}%
\thanks{$^{2}$These authors are with the Department of Mechanical Engineering, National University of Singapore, Singapore. Yutong Wang conducted the research during her visit to Carnegie Mellon University. \{yutong\_wang,e1280621\}@u.nus.edu,guillaume.sartoretti@nus.edu.sg}%
}

\begin{document}

\newcommand{\jh}[1]{{ \color{blue}[jh: #1]}}

\maketitle
\thispagestyle{empty}
\pagestyle{empty}

%%%%%%%%%%%%%%%%%%%%%%%%%%%%%%%%%%%%%%%%%%%%%%%%%%%%%%%%%%%%%%%%%%%%%%%%%%%%%%%%
\begin{abstract}

Lifelong Multi-Agent Path Finding (LMAPF) repeatedly finds collision-free paths for multiple agents that are continually assigned new goals when they reach current ones. Recently, this field has embraced learning-based methods, which reactively generate single-step actions based on individual local observations. However, it is still challenging for them to match the performance of the best search-based algorithms, especially in large-scale settings. This work proposes an imitation-learning-based LMAPF solver that introduces a novel communication module as well as systematic single-step collision resolution and global guidance techniques. Our proposed solver, Scalable Imitation Learning for LMAPF (SILLM), inherits the fast reasoning speed of learning-based methods and the high solution quality of search-based methods with the help of modern GPUs. Across six large-scale maps with up to 10,000 agents and varying obstacle structures, SILLM surpasses the best learning- and search-based baselines, achieving average throughput improvements of 137.7\% and 16.0\%, respectively. Furthermore, SILLM also beats the winning solution of the 2023 League of Robot Runners, an international LMAPF competition.
Finally, we validated SILLM with 10 real robots and 100 virtual robots in a mock warehouse environment.

\end{abstract}

%%%%%%%%%%%%%%%%%%%%%%%%%%%%%%%%%%%%%%%%%%%%%%%%%%%%%%%%%%%%%%%%%%%%%%%%%%%%%%%%
\section{Introduction}

Multi-Agent Path Finding (MAPF) \cite{SternSoCS19} is the problem of finding collision-free paths on a given graph for a set of agents, each assigned a start and goal location. Lifelong MAPF (LMAPF) \cite{ma2017lifelong} extends MAPF by continually assigning new goals to agents that reach their current ones. The main target of LMAPF is to maximize the throughput, which is defined as the average number of goals reached by all agents per timestep. LMAPF has wide applications in the real world, such as automated warehouses, traffic systems, and virtual games. For example, Amazon fulfillment centers were reported to have more than 4,000 robots deployed in 2022 \cite{Brown2023amazonrobot}. With growing demands in our daily lives, even larger autonomous systems are expected to be deployed in the near future. Therefore, more and more scalable search-based solvers have been developed for LMAPF in recent years \cite{li2021lifelong,okumura2022priority,chen2024traffic,Jiang2024Competition}, with state-of-the-art ones \cite{chen2024traffic,Jiang2024Competition} capable of scaling to thousands of agents.

Since learning-based solvers are expected to be more decentralized and scalable than search-based ones, numerous studies on learning have been conducted following the work PRIMAL~\cite{damani2021primal} since 2019.\footnote{Since solvers for MAPF can be easily adapted for LMAPF, we do not differentiate between MAPF and LMAPF in the rest of this paper, unless necessary.} 
% The success of deep learning in many other fields, such as computer vision and natural language, has also led researchers to investigate the potential of applying learning techniques in (L)MAPF.
% The mainstream following the pioneering work, PRIMAL~\cite{damani2021primal}, is to learn a neural policy that directly maps an agent's local View to an action. Such reactive planning can be fast with the help of modern GPUs.
However, most of them have only been tested on small-scale instances involving tens to hundreds of agents so far \cite{damani2021primal,li2022pico,wang2023scrimp,skrynnik2024decentralized,skrynnik2024learn}. 
%One potential reason is that directly exploring the large joint spaces of agents is hard. Many works need to imitate optimal search-based solutions, which can only be obtained in small instances. However, the generalization from small ones to large ones is non-trivial.

% Furthermore, many learning works illustrate their superior performance to search-based methods by comparing themselves with optimal or bounded suboptimal search algorithms, such as CBS~\cite{Sharon2015cbs} and ODrM*~\cite{wagner2015subdimensional}. 
% %However, these optimal search algorithms perform worse in practice primarily because of their long planning time. 
% Such comparisons are outdated since recent unbounded-suboptimal search-based solvers, such as Priority Inheritance with Backtracking (PIBT) \cite{okumura2022priority}, can quickly output relatively good solutions for large instances with thousands of agents. A very recent paper~\cite{skrynnik2024learn} shows that their learning-based LMAPF solver, Follower, can achieve higher throughput than PIBT, but we can also find that many previous state-of-the-art learning-based works, such as PICO~\cite{li2022pico} or SCRIMP~\cite{wang2023scrimp} actually fail to beat PIBT. Compared to RHCR~\cite{li2021lifelong}, Follower is faster but has 35\% lower throughput. Clearly, the quality of learning-based approaches can be improved further.
\begin{figure}[tb]
    \centering
    \includegraphics[width=0.99\linewidth]{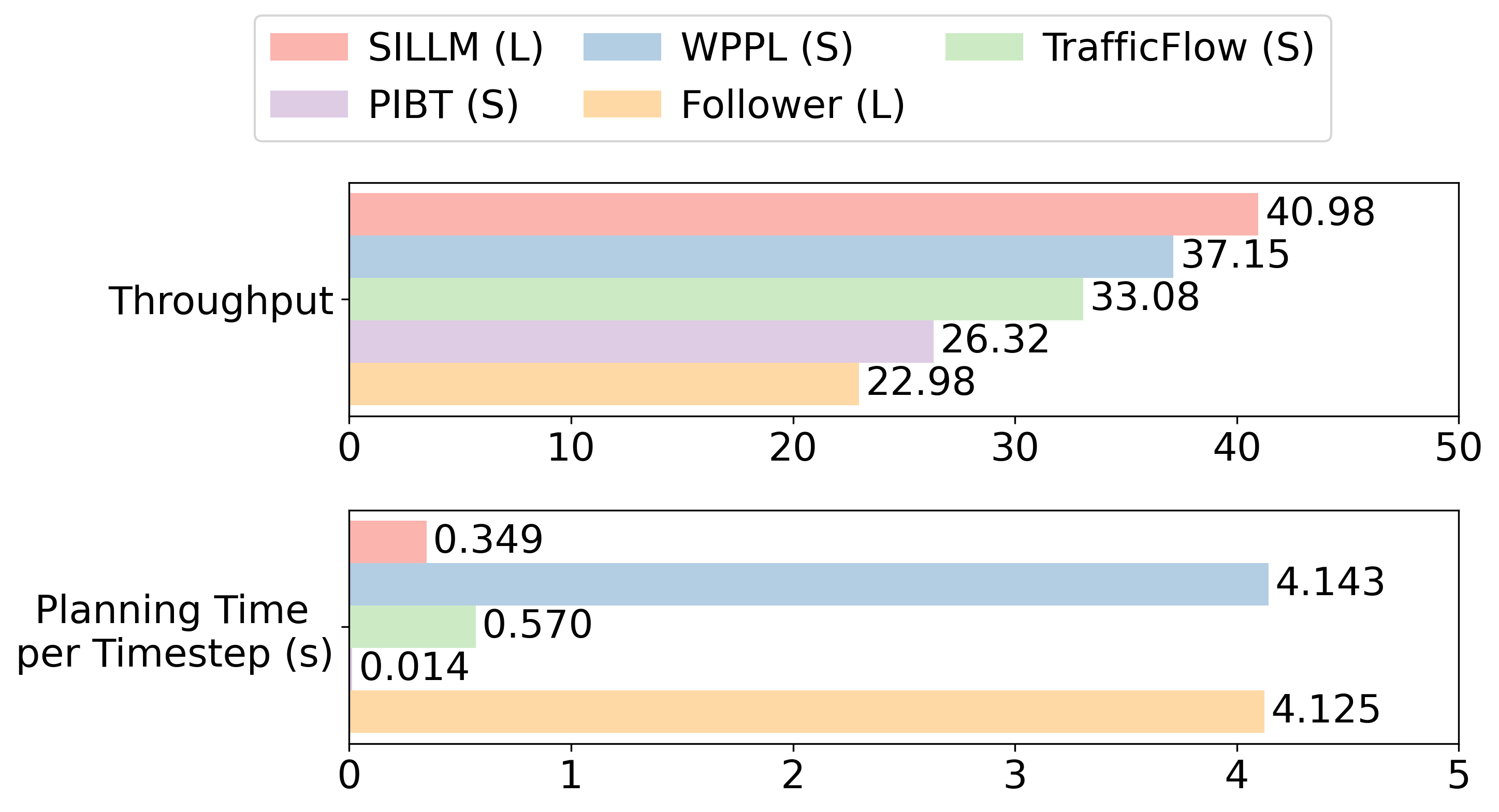}
    \caption{Comparison of mean throughput and mean planning time per step between our solver, SILLM,  with other state-of-the-art search- and learning-based solvers on 6 maps with 10,000 agents. The (L) and (S) in the legend denote learning-based and search-based solvers. Details are given in \Cref{tab:main}.}
    \label{fig: bar_compare}
    \vspace{-0.3cm}
\end{figure}

Additionally, most learning papers emphasize the scalability of their solvers compared to optimal or bounded-suboptimal search-based solvers \cite{Sharon2015cbs,ferner2013odrm}, often showing positive results. This is mainly because these search-based solvers struggle with computational complexity, as solving MAPF optimally is NP-hard \cite{yu2013structure}.
However, when compared to scalable suboptimal search-based solvers like PIBT \cite{okumura19pibt,okumura2022priority} (first introduced in 2019), most state-of-the-art learning-based solvers, such as PICO~\cite{li2022pico} or SCRIMP~\cite{wang2023scrimp}, would actually fail to beat them. Only very recently has a learning method Follower \cite{skrynnik2024learn} been shown to outperform PIBT in throughput on small maps with up to 192 agents.
% In addition, most learning papers focus on their solvers' scalability compared to optimal or bounded-suboptimal search-based solvers \cite{Sharon2015cbs,ferner2013odrm} and show positive results primarily because these optimal and bounded-suboptimal search-based solvers suffer from computational complexity (as MAPF is NP-hard to solve optimally). 
% A very recent learning paper~\cite{skrynnik2024learn} compares their solver, Follower, with PIBT and illustrates better performance, but we can also find that previous state-of-the-art learning-based works, such as PICO~\cite{li2022pico} or SCRIMP~\cite{wang2023scrimp}, actually fail to beat PIBT. %Compared to RHCR~\cite{li2021lifelong}, Follower is faster but has a 35\% lower throughput. Clearly, the quality of learning-based approaches can be improved further.
Indeed, recent state-of-the-art search-based solvers built on top of PIBT, such as TrafficFlow \cite{chen2024traffic} and WPPL \cite{Jiang2024Competition}, can act as even stronger baselines. Thus, our objective is to not only outperform existing learning-based methods but outperform existing search-based methods as well.
% It would be fairer if we could illustrate the advantage of learning over them.

In this paper, we propose \underline{S}calable \underline{I}mitation \underline{L}earning for \underline{LM}APF (SILLM), a learning-based solver that can manage up to $10,000$ agents.
Unlike prior works, we imitate a state-of-the-art scalable LMAPF solver on scenarios consisting of hundreds of agents~\cite{li2021anytime,Jiang2024Competition}.
We further design a novel communication architecture, the \underline{S}patially 
\underline{S}ensitive \underline{C}ommunication (SSC) module that emphasizes precise spatial reasoning to better learn highly cooperative actions seen in efficient LMAPF solutions.
Finally, we integrate heuristic search improvements in safeguarding single-step collisions~\cite{veerapaneni2024improving} and global guidance heuristics~\cite{Jiang2024Competition,chen2024traffic}.

Our experimental results show that SILLM outperforms state-of-the-art learning- and search-based solvers on six large maps from common benchmarks with up to 10,000 agents and varying obstacle structures, achieving average throughput improvements of 137.7\% and 16.0\%, respectively.
We further demonstrate that SILLM is even able to outperform WPPL \cite{Jiang2024Competition}, the winning solution of the 2023 League of Robot Runners \cite{chan2024league}, specially designed to solve large-scale LMAPF problems. Notably, with the aid of GPUs, SILLM takes less than 1 second to plan 10,000 agents at each timestep. \Cref{fig: bar_compare} illustrates a simple comparison of average throughput and planning time.
Furthermore, we validated SILLM with 10 real robots and 100 virtual robots in a mock warehouse environment, further showcasing its potential for real-world applications. Our work highlights the effectiveness of learning-based methods for large-scale LMAPF instances and offers a comprehensive framework to unlock the full power of learning in future research.
% Our experiment results show that SILLM surpasses both the best learning- and search-based baselines, achieving average throughput improvements of 137.7\% and 16.0\%, respectively, across six large maps with 10,000 agents and varying obstacle structures. We further show that SILLM can even beat WPPL \cite{Jiang2024Competition}, the winning solution of the 2023 League of Robot Runners \cite{chan2024league}, an international competition sponsored by Amazon Robotics that focuses on solving challenging LMAPF instances with up to 10,000 agents. A simple illustration of performance comparison is shown in \Cref{fig: bar_compare}. Importantly, with the aid of GPUs, SILLM takes less than 1 second to plan for 10,000 agents at each timestep.
% We finally validated SILLM with $10$ real robots and $100$ virtual robots, respectively, in a mock warehouse environment.

% Our work validates the potential of learning-based methods for large-scale LMAPF instances and provides a holistic framework for future work to unleash the power of learning. 

% ? what can we claim 
% review previous learning works
% 

\section{Background}
The LMAPF problem includes a $4$-neighbor grid graph $G = (V, E)$ and a set of $n$ agents $A = \{a_1, \ldots, a_n\}$, each with a unique start location.
The vertices $V$ of the graph $G$ correspond to locations (namely, unblocked grid cells), and the edges $E$ correspond to connections between neighboring locations. %In this paper, we only consider $4$-connected grid graphs. 
Time is discretized into timesteps. At each timestep, an agent can move to an adjacent location by one of four move actions (up, down, left, right) or wait at its current location. During execution, we disallow vertex collisions and edge collisions. A vertex collision occurs when two agents occupy the same location at the same timestep. An edge collision occurs when two agents traverse the same edge in opposite directions at the same timestep. 

%In contrast to standard MAPF, where each agent is assigned a single unique goal, 
LMAPF requires repeated planning as agents are continuously assigned new goals. We assume that an external task assigner handles goal assignments, with each agent knowing its next goal only upon reaching its current one. An LMAPF instance runs for a predetermined number of timesteps. The objective is to plan collision-free paths for all agents while maximizing throughput, i.e., the average number of goals reached per timestep.

\section{Related Work}
PRIMAL~\cite{sartoretti2019primal} is the pioneering work that adopts learning to solve the MAPF problem, which exploits the homogeneity in agents to train a shared decentralized policy.
Subsequent research has focused on improving it from four main directions: enabling communication between agents~\cite{ma2021DHC,wang2023scrimp}, incorporating global guidance into agents' field of view (FoV)~\cite{wang2020G2RL,skrynnik2024learn}, enhancing collision resolution~\cite{wang2023scrimp}, and imitating search-based MAPF algorithms \cite{damani2021primal,li2021magat}.
In~\Cref{table:review}, we summarize the learning-based solvers proposed in recent years that are commonly used as baselines, and in each subsection of \Cref{s: methods}, we explain the improvements we made to each technique.

In addition to learning-based solvers, we are also interested in the recent progress in search-based methods. RHCR \cite{li2021lifelong} is one of the most widely used baselines for LMAPF, which introduces the planning window to reduce the computation. It has a high solution quality but is still relatively slow. As a result, it can only manage hundreds of agents. On the other hand, PIBT \cite{okumura2022priority} is a greedy single-step planner, which is very fast and scalable. It can coordinate thousands of agents but has a relatively low solution quality. TrafficFlow \cite{chen2024traffic} incorporates traffic information to help PIBT avoid congestion and, thus, improves its solution quality. As a competitive approach, WPPL \cite{Jiang2024Competition} is the winning solution of the 2023 League of Robot Runner Competition \cite{chan2024league}, which exploits PIBT to generate an initial windowed plan and then applies windowed MAPF-LNS \cite{li2021anytime} to refine the plan.

% % \section{Related Work}
% PRIMAL\cite{sartoretti2019primal} is the pioneering work that adopts learning-based approaches for the MAPF problem, which proposed to leverage homogeneity in MAPF to learn a decentralized policy which iterativeky do single-step dicision makeing shared by all agents.
% Subsequent research has primarily focused on improving learning in MAPF from four aspects, including add comunication between agents, incooperate global guidance culated by serach-based algrithms, imporve collsion resolving efficiencym and imitate to a search-based MAPF algrithms. 
% We summarize representative learning-based works which propose in recent years and is the common baseline in the learning-based MAPF arear in table \Cref{table:review} and analyze the improvemt teniques in each subsection of \Cref{s: methods} Methods.

\begin{table*}[t]
\setlength{\tabcolsep}{3pt}
\centering
\caption{
Overview of learning-based solvers.
The ``Max \#Agents" is based on experiments reported in the papers.
Some methods allow communication between agents: ``ABC" denotes attention-based communication, and ``SSC" denotes our spatially sensitive communication.
All methods use agents' FoV as inputs, while some also incorporate global guidance, such as a path calculated by A* (Path) or binary features indicating whether a movement brings the agent closer to its goal (Movements).
Our method, SILLM, uses three types of global guidance: Backward Dijkstra (BD), Static Guidance (SG), and Dynamic Guidance (DG), which are explained in \Cref{ss: heuristics}.
At each timestep, collisions are resolved by either keeping agents in their previous locations (Freeze), allowing them to reselect new actions (Reselect), or applying Collision Shield PIBT (CS-PIBT) \cite{veerapaneni2024improving}.
All methods are trained via imitation learning (IL), reinforcement learning (RL), or both (Mixed). IL uses search-based solvers as imitation objectives. 
% Each method targets either one-shot MAPF, LMAPF, or both.
% The "Map type" and "Maximum Number of Agents" (Max \#Agents) are based on the experiments reported in the original paper.
% "Map type" refers to the characteristics of obstacle distribution in the tested maps, where random obstacle maps with different obstacle densities are considered to have the same obstacle distribution characteristics.
% All methods use agents' local views as inputs, while some also incorporate global guidance,such as path caluated by A* or weighte A* and actions by Whether the Move Brings Agent Closer to Goalaccording to and Backward Dijkstra (BD).
% All methods use agents' local views as inputs, while some also incorporate global guidance computed by search-based algorithms like A* and Backward Dijkstra (BD).
% Transferring models without retrain to previously unseen scenarios often results in performance degradation.
}
\label{table:review}
\resizebox{1\textwidth}{!}{
\begin{tabular}{c|ccccccccccc}
\toprule
\rowcolor[HTML]{FFFFFF} 
 & \begin{tabular}[c]{@{}c@{}}PRIMAL\\2019~\cite{sartoretti2019primal}\end{tabular} & \begin{tabular}[c]{@{}c@{}}MAPPER\\~2020\cite{liu2020mapper}\end{tabular} & \begin{tabular}[c]{@{}c@{}}PRIMAL2\\2021~\cite{damani2021primal}\end{tabular} & \begin{tabular}[c]{@{}c@{}}MAGAT\\2021~\cite{li2021magat}\end{tabular} & \begin{tabular}[c]{@{}c@{}}DHC\\2021~\cite{ma2021DHC}\end{tabular} & \begin{tabular}[c]{@{}c@{}}DCC\\2021~\cite{ma2021learning}\end{tabular} & \begin{tabular}[c]{@{}c@{}}SACHA\\2023~\cite{lin2023sacha}\end{tabular} & \begin{tabular}[c]{@{}c@{}}RDE\\2023~\cite{gao2023rde}\end{tabular}  & \begin{tabular}[c]{@{}c@{}}SCRIMP\\2023~\cite{wang2023scrimp}\end{tabular}& \begin{tabular}[c]{@{}c@{}}FOLLOWER\\2024~\cite{skrynnik2024learn}\end{tabular} & \begin{tabular}[c]{@{}c@{}}SILLM\\(Ours)\\   \end{tabular} \\ 

\midrule
\rowcolor[HTML]{EFEFEF} 
% Task & MAPF & MAPF & (L)MAPF & MAPF & MAPF & MAPF & MAPF & MAPF & MAPF & LMAPF & LMAPF \\
% \rowcolor[HTML]{EFEFEF} 
% Map Type & Random & Mixed Obstacles & Maze, Warehouse & Random & Random & Warehouse & Random & \begin{tabular}[c]{@{}c@{}}Random, Maze, \\ Game, City, Warehouse\end{tabular} & \begin{tabular}[c]{@{}c@{}}Random, Maze, Game, City,\\  Warehouse, Sortation\end{tabular} \\
% Map Types & 1 & 1 & 2 & 1 & 1 & 1 & 3 & 1 & 1 & 5 & 5 \\
% \rowcolor[HTML]{FFFFFF} 
\rowcolor[HTML]{EFEFEF} 
Max \#Agents & 1,024 & 150 & 2,048 & 1,000 & 64 & 128 &64 & 70 & 128 & 256 & 10,000 \\ 
% Max Map Size \\
% \rowcolor[HTML]{FFFFFF} 
Communication & None & None & None & ABC & ABC & ABC & ABC & ABC & ABC & None & SSC \\
% \rowcolor[HTML]{EFEFEF}
\rowcolor[HTML]{EFEFEF} 
Global Guidance & None & Path & Path &None & Movements & Movements & Movements & Movements & Movements &  Path & BD,SG,DG\\
% Global Guidance &  & BD & BD &  & BD & BD & BD & DG & BD,SG,DG\\
% Guiding Info &  & \begin{tabular}[c]{@{}c@{}}Individual Path \\ Calculated by A*\end{tabular} & \begin{tabular}[c]{@{}c@{}}Agents' Next Three \\ Positions Calculated by A*\end{tabular} &  & \begin{tabular}[c]{@{}c@{}}Whether the Move Brings \\ Agent Closer to Goal\end{tabular} & \begin{tabular}[c]{@{}c@{}}Whether the Move Brings \\ Agent Closer to Goal\end{tabular} & \begin{tabular}[c]{@{}c@{}}Whether the Move Brings \\ Agent Closer to Goal\end{tabular} & \begin{tabular}[c]{@{}c@{}}Individual Path \\ Calculated by Weighted A*\end{tabular} & \begin{tabular}[c]{@{}c@{}}Normalized \\ Heuristic Values\end{tabular} \\
% \rowcolor[HTML]{FFFFFF} 
Collision Resolution & Freeze & Freeze & Freeze & Freeze & Freeze & Freeze & Freeze & Freeze & Reselect & Freeze & CS-PIBT \\
\rowcolor[HTML]{EFEFEF} 
Training Approach & Mixed & RL & Mixed & IL & RL& RL & RL & RL & Mixed & RL & IL \\
% RL & \checkmark & \checkmark & \checkmark &  & \checkmark & \checkmark & \checkmark & \checkmark &  \\
% \rowcolor[HTML]{EFEFEF} 
% IL & \checkmark & & \checkmark & \checkmark &  &  & \checkmark &  & \checkmark \\
% \rowcolor[HTML]{EFEFEF} 
Imitation Objective & ODrM* & None & ODrM* & ECBS & None& None &None & None & ODrM* & None & W-MAPF-LNS \\

\bottomrule
\end{tabular}
}
\end{table*}

\section{Methods}
\label{s: methods}

In this section, we first describe the network structure of SILLM, including the Spatially Sensitive Communication (SSC) module, in \Cref{ss: spatially sensitive communication}. Then, we discuss three types of global guidance in \Cref{ss: heuristics}. Finally, we introduce the inference and training procedure in \Cref{ss: CS-PIBT,ss: Windowed MAPF-LNS}.

\subsection{Neural Policy with Spatially Sensitive Communication}
\label{ss: spatially sensitive communication}

Communication is an important aspect of neural network design in multi-agent systems. As shown in the second row of \Cref{table:review}, existing works mostly adopted attention-based communication (ABC) to aggregate information from neighboring agents, which only implicitly reasons with the spatial information. However, we argue that precise spatial information benefits local collision avoidance, as similar designs, such as the conflict avoidance table \cite{standley2010finding}, are frequently applied in search-based solvers. Therefore, we propose a spatial sensitive communication (SSC) module that explicitly preserves the spatial relationship between agents when aggregating information.

\begin{figure}[tb]
    \centering
    \includegraphics[width=1.0\linewidth]{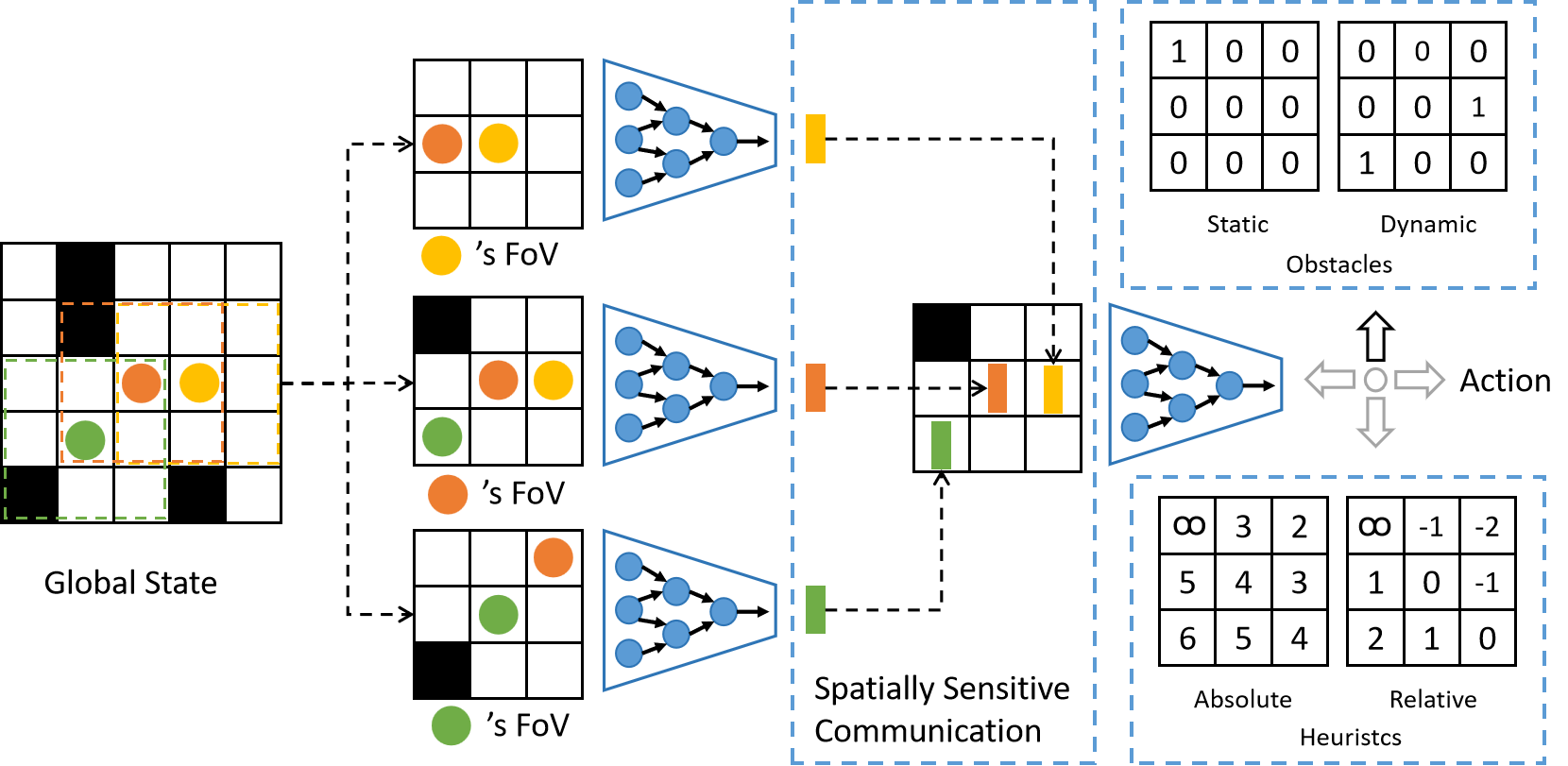}
    \caption{Core network structure. The global state has all static obstacles (black squares) and agents (colored circles). As an example, an agent's FoV is of size $3\times3$. The orange agent's unnormalized obstacle and heuristic feature maps are shown in the right upper and bottom corners. %The details are in the \cref{ss: spatially sensitive communication}.
    %The first CNN extracts feature vectors for all agents based on their local observations. These feature vectors are then inserted into the orange agent's local view according to each agent's relative position in the Spatially Sensitive Communication module. Finally, another CNN outputs the final action based on the new local observation built from the feature vectors.
    }
    \label{fig: neural policy}
     \vspace{-0.3cm}
\end{figure}
\Cref{fig: neural policy} illustrates the core structure of our neural network. The neural network comprises two Convolutional Neural Network (CNN) modules and a Spatially Sensitive Communication (SSC) module. Since all agents are homogeneous in LMAPF, they share the same network weights.

Each agent $a_i$ has a local FoV of size $V_h \times V_w$ (set to $V_h=V_w=11$ in our experiments), centered at the agent's position, with a total of $5$ feature channels. These channels are divided into two overlapping groups $o_1^i$ and $o_2^i$. $o_1^i$, with a size of $4 \times V_h \times V_w$, consists of four channels representing the locations of static obstacles and dynamic obstacles (i.e., other agents), as well as absolute and relative heuristic values (will be explained in \Cref{ss: heuristics}). $o_2^i$, with a size of $3 \times V_h \times V_w$, consists of three channels representing the locations of static and dynamic obstacles and the goal of agent $a_i$.

We first use a CNN module to convert $o_1^i$ into a $32$-channel feature vector $f_i$. Then, $a_i$ gathers the feature vectors of all neighboring agents within its local FoV through communication.\footnote{The communication range can be smaller than the FoV, but for simplicity, we set them to be the same in the experiments.} Subsequently, the SSC module generates a matrix $o_3^i$ of shape $32 \times V_h \times V_w$, filled with $-1$, and inserts the gathered feature vectors into the corresponding agents' relative positions in $a_i$'s FoV (illustrated in \Cref{fig: neural policy}). $o_3^i$ is then element-wise added to a matrix of the same shape, obtained by processing $o_2^i$ through a Conv2d layer with a kernel size of 1. Finally, the resulting matrix undergoes further feature extraction through another CNN module and is decoded into the probabilities of five actions.

\subsection{Providing Global Guidance with Heuristics}
\label{ss: heuristics}

% Learning-based solvers require global guiding information to indicate the goal location, as summarised in the third row of \Cref{table:review}. The naive approach encodes the coordination, direction, or projection of the goal location~\cite{sartoretti2019primal,li2021magat}. However, they are more suitable for map exploration but less informative for applications with known maps. Recent works mostly encode information based on the shortest path to the goal. 
Many learning-based works incorporate global guidance to help agents move toward their goals, as summarized in the third row of \Cref{table:review}.
Some works try to follow a specific shortest path \cite{wang2020G2RL,liu2020mapper}. However, since the shortest path might not be unique, other works encode the shortest distance from every location in FOV to the goal~\cite{ma2021DHC,wang2023scrimp,gao2023rde,lin2023sacha,skrynnik2024decentralized}.  A recent paper, Follower~\cite{skrynnik2024learn}, tries to follow a shortest path that considers local traffic and is replanned at each timestep. However, no existing work reduces global traffic, which is important for systems with large agent numbers.

We systematically study three types of global guidance represented as heuristics. The first type of heuristics is the aforementioned \textbf{Backward Dijkstra (BD)} heuristics applied in previous learning works, which tell agents the shortest distances to their goals. Further, we evaluate two other heuristics that were not applied in the previous learning works but effectively reduce global traffic.

The second type of heuristics is still Backward Dijkstra heuristics but based on specially designed edge costs that reduce traffic offline. We call it \textbf{Static Guidance (SG)}. SG heuristics encourage agents to move in the same direction, avoiding head-on collisions. Specifically, we adopt the classic crisscross highways~\cite{lironPhDthesis}, where the encouraged directions are alternating row by row and column by column. In practice, the default cost of an edge is 2. The cost of an edge in the encouraged directions is 1, and the cost in the discouraged directions is 100,000 for warehouse or sortation maps and 3 for other maps.
% In this paper, we adopt a classical design, the crisscross highway~\cite{lironPhDthesis}, where edge costs are set to encourage agents to move in a single direction at each row/column and alternate the encouraging directions row/column by row/column. For example, in the first row, the edge cost of moving left is designed to be less than the edge cost of moving right. Then, agents will tend to move left during planning. For the second row, the design is exactly opposite, and agents will tend to move to the right. Since fewer agents move in the opposite directions in a row/column, the traffic is largely reduced. 
More advanced edge cost designs are also possible. For example, GGO \cite{zhang2024ggo} proposes an automatic way to optimize edge costs to maximize throughput. % but it cannot be easily applied to large maps due to the heavy computation.

The last type of heuristics is \textbf{Dynamic Guidance (DG)}, which further encodes dynamic traffic information,  encouraging agents to move along short paths while avoiding congestion. We adopt the implementation in TrafficFlow \cite{chen2024traffic}, which plans a guide path for each agent and then asks agents to follow their guide paths as much as possible. When planning a guide path for an agent, TrafficFlow first counts the global traffic information based on the guide paths of other agents. Specifically, it counts the number of other agents visiting each location and traversing each edge along their guide paths. Then, dynamic edge costs are defined by handcrafted equations to punish the agent for moving to frequently visited locations and moving in the opposite directions of frequently visited edges. Therefore, the planned guide path could avoid potentially congested regions.\footnote{Readers interested in the implementation details are encouraged to refer to the original paper. We use the best variant in the paper.}

Given the guide path of an agent $a_i$ whose current goal is $g_i$, the heuristic value of location $v$ for $a_i$ is defined as 
\begin{align*}
    h(v)= SPCost(v, v') + GPCost(v', g_i), \label{eq: hTF}
\end{align*}
where $v'$ denotes the closest location to $v$ in the guide path, $SPCost(v, v')$ represents the shortest path cost from $v$ to $v'$, and $GPCost(v', g_i)$ is the remaining path cost from $v'$ along the guide path to goal $g_i$. %Notably, $SPCost$ can be easily computed by a backward A* search starting from all the locations in the guide path, while $GPCost$ can be directly obtained by summing the edge costs along the guide path.

We encode these three different types of heuristics in a unified way into the observation. Specifically, we denote the heuristic value at location $v$ as $h(v)$ and encode the heuristic values in the FoV by a $2$-channel 2D feature map that has the same size as the FoV. The first channel is the absolute heuristic value normalized by the map size, i.e., the feature value at location $v_i$ is $h(v_i)/(M_h+M_w)$, where $M_h$ and $M_w$ are the height and width of the map $G$. The second channel is the relative difference in heuristic values, calculated by subtracting the heuristic value at the center of the FoV and dividing the FoV size, i.e., the feature value at location $v_i$ is $(h(v_i)-h(v_c))/(V_h+V_w)$, where $V_h$ and $V_w$ are the height and width of the FoV, and $v_c$ is the center of the FoV. % Even though we explicitly add value normalization, we apply a Batch Normalization to the input to accommodate the potential value range problem. 
% % We discuss how $h(p)$ is consistently incorporated into the optimization objective of W-MAPF-LNS in \Cref{ss: Windowed MAPF-LNS}.

% need an example for reference path based heuristics.

\subsection{Safeguarding Single-Step Execution with CS-PIBT}
\label{ss: CS-PIBT}
% We adapt the CS-PIBT with a little modification.

Our neural policy has one remaining issue during inference: the generated actions may still contain collisions as they are independently sampled by agents. Most earlier works address it by ``freezing" the agents' movements that potentially lead to collisions by replacing their original actions with wait actions \cite{sartoretti2019primal}, as shown in the fourth row of \Cref{table:review}. However, this approach is inefficient as many unnecessary wait actions are introduced. %, especially in congested scenarios where agents' movements are highly dependent. 
It can also lead to deadlocks when agents repeatedly select the same conflicting actions. To mitigate this issue, SCRIMP \cite{wang2023scrimp} allows agents to reselect actions based on their action probabilities to avoid the original conflicts, but it has trouble scaling to many agents. 
% This reselection strategy significantly improves the performance compared to ``freeze" but requires more computation time since agents may need to reselect actions multiple times to resolve all collisions.
Collision Shield PIBT (CS-PIBT) \cite{veerapaneni2024improving} proposes to use PIBT \cite{okumura2022priority} to resolve collisions.

In PIBT, each agent is assigned a priority. At each step, every agent ranks its actions in ascending order of the shortest distance from the resulting location to its goal (i.e., the BD heuristic). An agent always takes its highest-ranked action that does not collide with any higher-priority agent. If no collision-free actions exist for a low-priority agent, PIBT triggers a backtracking process, forcing the higher-priority agent to take its next best action until all agents can take collision-free actions. Given an agent's action probabilities from the neural network, CS-PIBT converts them into a ranking (by biased sampling) and runs PIBT with this action preference to get 1-step collision-free actions.

% Notably, we can flexibly replace the preference of actions by other heuristics \cite{veerapaneni2024improving,chen2024traffic} in PIBT. 

%Notably, other heuristic values can flexibly replace the shortest distances to goals in this process \cite{veerapaneni2024improving,chen2024traffic}. %For example, CS-PIBT \cite{veerapaneni2024improving} studies different ways of combining a neural policy's output and the shortest distances. 

We directly applied this idea in our work with slight modifications; unlike CS-PIBT, which safeguards the inference of a policy trained by other methods, this work holistically considers training and inference. We adopt a simplified variant of CS-PIBT, which always prioritizes the learned action output by the neural policy and then ranks other actions as the original PIBT. If the learned actions are collision-free, then CS-PIBT will not change them. We still call this safeguarding procedure CS-PIBT but name the combination of the neural policy and CS-PIBT as Learnable PIBT (L-PIBT) from the perspective of learning. 

%Such a simple design has the advantage that it can be seamlessly plugged into any training procedure, including reinforcement learning, by replacing the original environment with a new one that automatically applies CS-PIBT to the actions before simulating the environment step. Then, the learning algorithms do not need to worry about safety issues and can train the neural policy as usual.

%In our case, we rank actions preferring less $h'$ in the following.
% \begin{align*}
%     h'(v_n)=
% \begin{cases}
% -1, \text{if $v_n$ is the next location by taking action $\pi(v)$}\\
% h(v_n), \text{otherwise} \\
% \end{cases}
% \end{align*}

\subsection{Imitation Learning From A Scalable Search-Based Solver}
\label{ss: Windowed MAPF-LNS}

\begin{figure}[tb]
    \centering
    \includegraphics[width=1.0\linewidth]{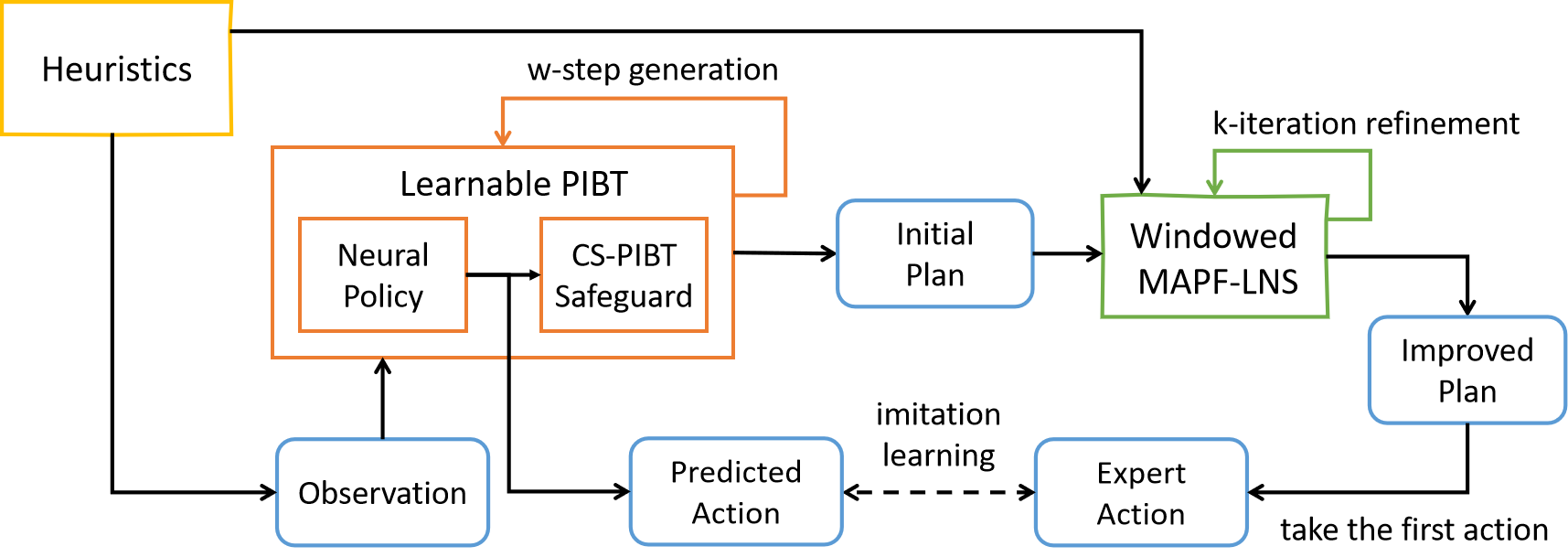}
    \caption{Data collection procedure in the \Cref{ss: Windowed MAPF-LNS}.
    %All the data are marked as blue boxes. The orange box is Learnable PIBT, composed of a neural policy and CS-PIBT. CS-PIBT is used to safeguard single-step execution. The green box is W-MAPF-LNS, which generates highly cooperative actions for the neural policy to imitate. The yellow box is Heuristics, which provides global guidance to agents to mitigate myopia of the neural policy.
    }
    \label{fig: training framework}
     \vspace{-0.3cm}
\end{figure}

We choose imitation learning because it is generally easier to learn team cooperation from mature search-based solvers, while reinforcement learning needs to explore the vast joint action space of multiple agents. A few earlier works have applied imitation learning, and they primarily focus on mimicking weak bounded-suboptimal algorithms, 
such as ODrM*~\cite{ferner2013odrm} or ECBS~\cite{barer2014suboptimal}, as shown in the last row of \Cref{table:review}. As a result, they are typically constrained to small-scale instances due to the heavy computation in data collection. For instance,  PRIMAL2 \cite{damani2021primal} and SCRIMP \cite{wang2023scrimp} %employed inflated ODrM* \cite{ferner2013odrm} and 
were trained on instances with $8$ agents, while MAGAT \cite{li2021magat} %utilized ECBS \cite{barer2014suboptimal} and 
was trained on instances with $10$ agents.
When applying these solvers to large-scale instances, the significant differences between the original training setup and the actual application scenarios often result in a substantial decline in performance. 

To address this issue, we directly imitate an anytime search-based algorithm, Windowed MAPF-LNS (W-MAPF-LNS), the central part of the winning solution WPPL~\cite{Jiang2024Competition}. It is unbounded suboptimal but scales well to large instances due to its planning window and anytime behavior. 

The data collection procedure for imitation learning is shown in \Cref{fig: training framework}. Given the observation at the current step $T$, Learnable PIBT is called $w$ times to generate the initial $w$-step paths for all agents. Then, we apply W-MAPF-LNS to refine their $w$-step paths for $k$ iterations. At each iteration, W-MAPF-LNS selects a small group of agents heuristically and tries to improve their $w$-step paths. Specifically, W-MAPF-LNS tries to optimize agents' $w$-step paths for an approximate objective:
\begin{align*}
    Obj=\sum_{i=1}^n(\sum_{t=T}^{T+w-1} Cost(v_t^i,v_{t+1}^i)+h(v^i_{T+w})),
    %\label{eq: obj}
\end{align*}
where $v_t^i$ is agent $a_i$'s location at step $t$. The $Cost$ function records the edge cost from one location to a neighbor, and $h(v^i_{T+w})$ is the heuristic value in \Cref{ss: heuristics}, which estimates the future cost from $v^i_{T+w}$ to agent $i$'s goal. Notably, the heuristics are consistently used in the observation and objective to make policy learning easy. Empirically, we set $w=15$ and $k=5,000$ so that the planning time at each step is less than $1$ second during data collection. 

Then, we collect the first actions in the refined $w$-step paths with the current observation for later supervised training of Learnable PIBT. Notably, the imitation learning procedure can be repeated iteratively in a self-bootstrapping manner. After an iteration of supervised training, we obtained a better Learnable PIBT, which could generate better initial $w$-step paths. Then, we can potentially obtain better improved $w$-step paths by W-MAPF-LNS, whose first actions are further used for the supervised learning of Learnable PIBT. Empirically, each iteration collects $15$ million action-observation pairs in $50$ episodes. We iterate the imitation 12 times and always select the best checkpoint in validation.

\begin{table}[tb]
\setlength{\tabcolsep}{3pt}
    \caption{Map visualization with details below. Sortation (top left) and Warehouse (bottom left) maps are from the LMAPF Competition \cite{chan2024league}, only showing the top left corners. Paris (middle) and Berlin (right) maps are from the MovingAI benchmark \cite{SternSoCS19}. Random1 and Random2 are randomly generated maps with 10\% and 20\% obstacle densities. We use the underlined characters as the abbreviation for each map. \#Locs is the number of unblocked locations, and 
    % \#Steps is the number of simulation steps.
    Agent density is defined as the number of agents divided by \#Locs.
    }
    \label{tab:instance}

% \begin{minipage}{0.5\textwidth}
%     \centering
%     \includegraphics[width=1.0\linewidth]{figures//main/maps.png}
%     % \caption{Top Left: Sortation. Bottom Left: Warehouse. Middle: Paris. Right: Berlin.}
%     \label{fig:enter-label}
% \end{minipage}
\centering
\begin{minipage}{0.155\textwidth}
    \centering
    \includegraphics[width=1.0\linewidth]{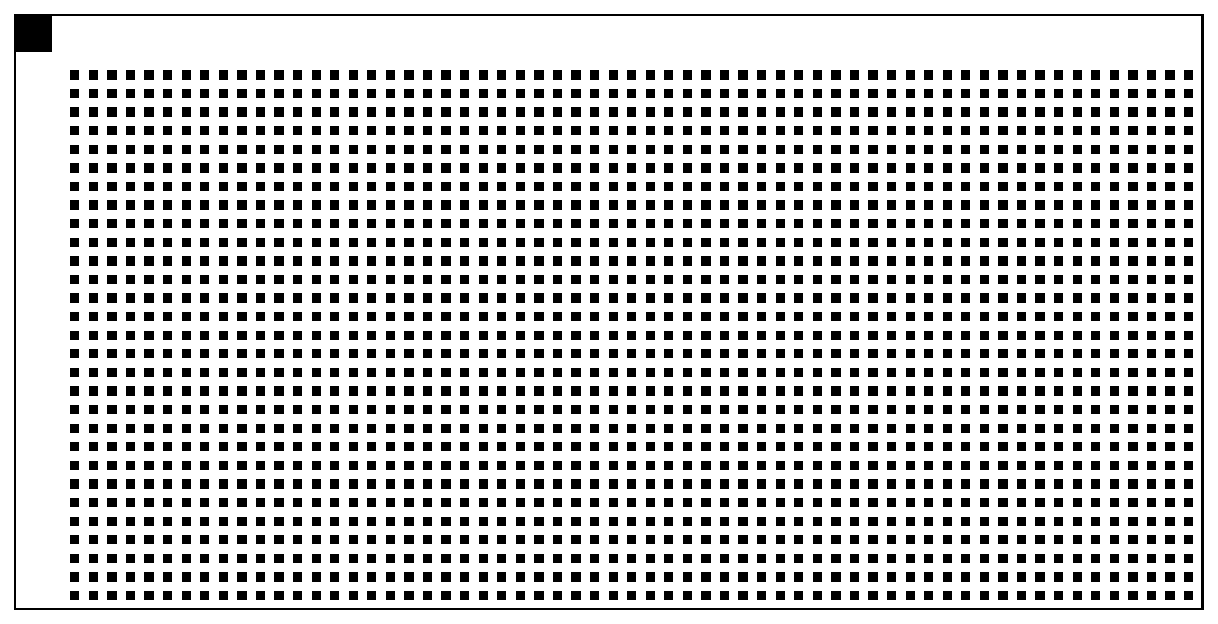}
    
    \includegraphics[width=1.0\linewidth]{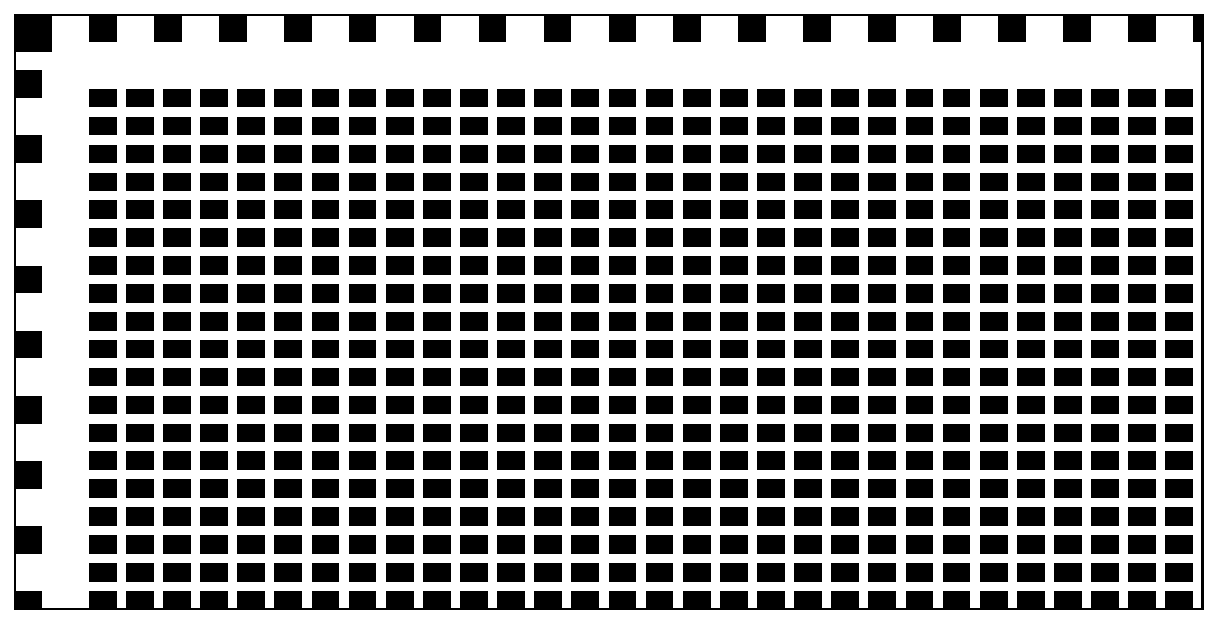}
\end{minipage}
\begin{minipage}{0.155\textwidth}
    \centering
    \includegraphics[width=1.0\linewidth]{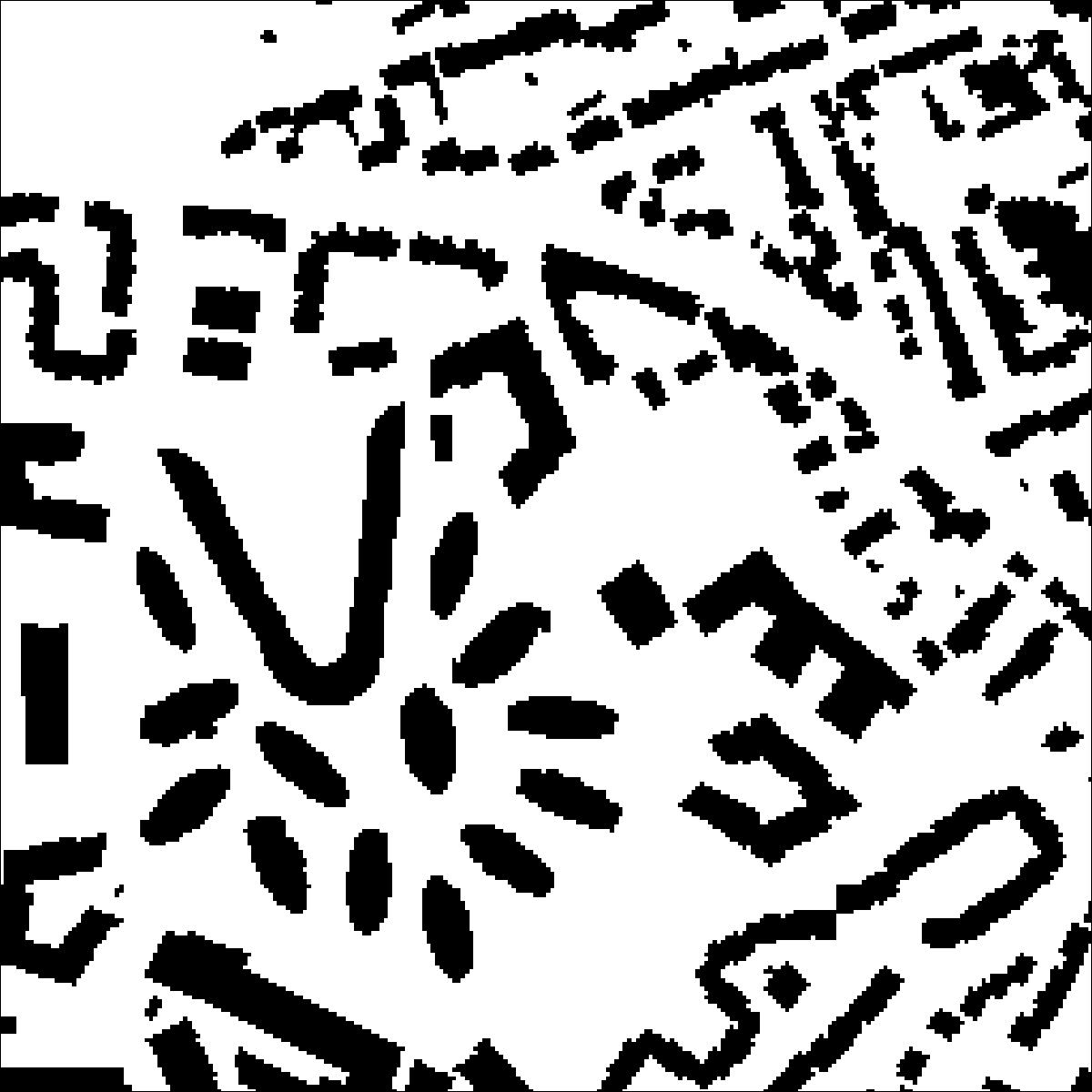}
\end{minipage}
\begin{minipage}{0.155\textwidth}
    \centering
    \includegraphics[width=1.0\linewidth]{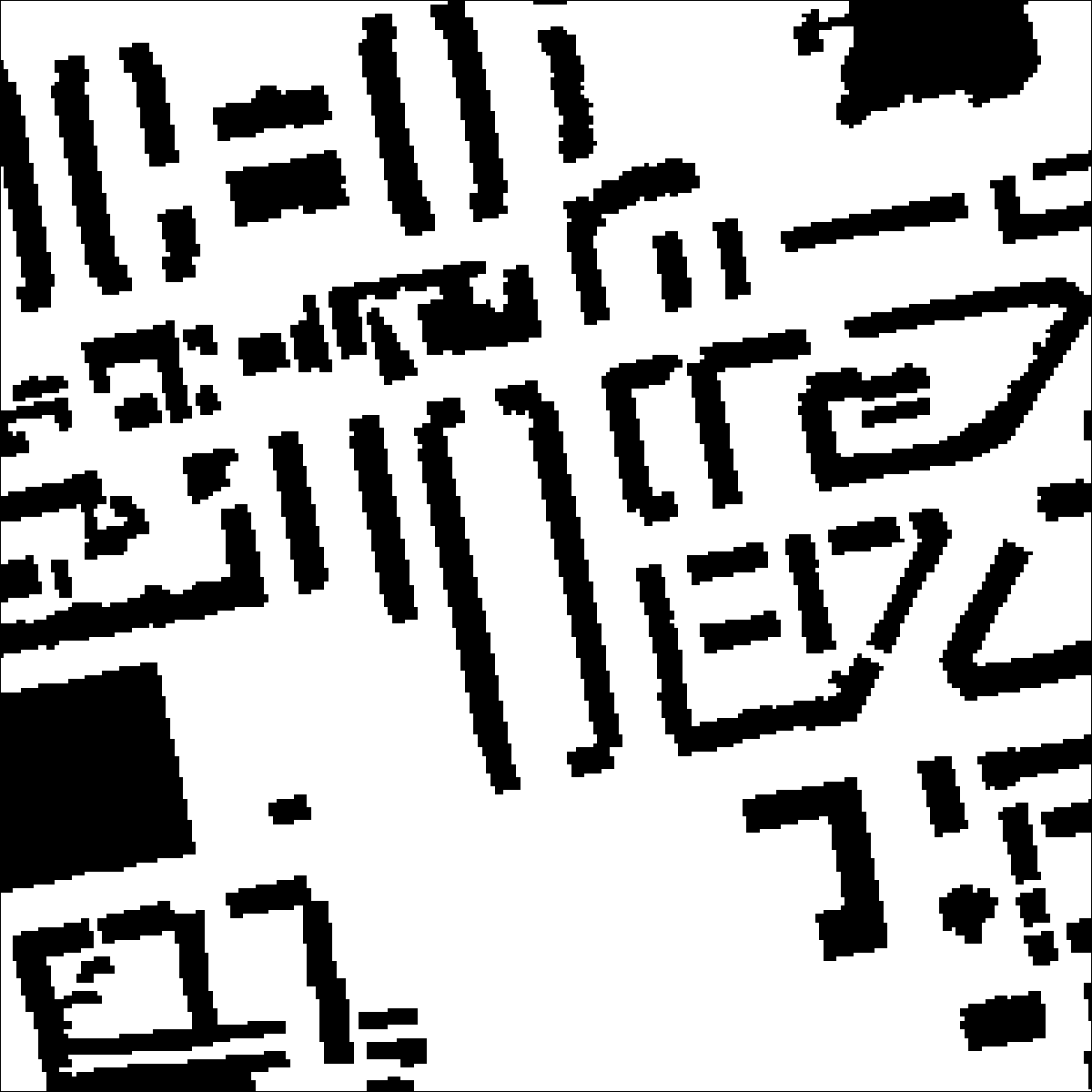}
\end{minipage}

\vspace{0.1cm}

\centering
\begin{minipage}{0.5\textwidth}
    \centering
    \resizebox{1.0\textwidth}{!}{
    \begin{tabular}{c|ccc|ccc}
    \toprule
        \multirow{2}{*}{Name}  & \multicolumn{3}{c|}{Small-Scale} & \multicolumn{3}{c}{Large-Scale}\\
        %\cmidrule{2-14}
          & Size & \#Locs & \makecell{Agent\\ Density} & Size & \#Locs & \makecell{Agent\\ Density} \\

        \midrule
        \underline{Sort}ation & 33*57 & 1,564 & 38.3\% & 140*500 & 54,320 & 18.4\% \\
        \underline{Ware}house & 33*57 & 1,277 & 47.0\% & 140*500 & 38,586 & 25.9\% \\
        \underline{Pari}s & 64*64 & 3,001 & 20.0\% & 256*256 & 47,096 & 21.2\% \\
        \underline{Berl}in & 64*64 & 2,875 & 20.9\% & 256*256 & 46,880 & 21.3\% \\
        \underline{Ran}dom\underline{1} & 64*64 & 3,670 & 16.3\% & 256*256 & 58,859 & 17.0\% \\
        \underline{Ran}dom\underline{2} & 64*64 & 3,221 & 18.6\% & 256*256 & 52,090 & 19.2\% \\

\bottomrule
    \end{tabular}
    }
    \end{minipage}
    \vspace{-0.3cm}
\end{table}

\begin{table*}[htb]
\setlength{\tabcolsep}{3pt}
    \caption{Comparison of different solvers. The left part is the result on downscaled small instances. The right part is the result on the original large instances. We evaluate each instance with $8$ runs of different starts and goals. Each column records the mean throughput with the standard deviation in the parentheses. The Time (in seconds) and Score refer to the average single-step planning time and the average score, respectively. ``-" indicates unavailable data due to the excessive planning time. We rank the solvers in descending order based on their average score on large-scale instances. The best throughput of each map is marked in bold. Notably, PIBT and TrafficFlow are exactly PIBT+BD and PIBT+DG in \Cref{fig: radar}. 
    }
    \label{tab:main}
    \centering
    \resizebox{1\textwidth}{!}{
    \begin{tabular}{c|cccccc|cc|cccccc|cc}
    % \toprule
    \toprule
        \multirow{2}{*}{Algorithm}  & \multicolumn{8}{c|}{Small maps with 600 agents} & \multicolumn{8}{c}{Large maps with 10,000 agents}\\
        %\cmidrule{2-14}
         & Sort & Ware & Pari & Berl & Ran1 & Ran2 & Time & Score & Sort & Ware & Pari & Berl & Ran1 & Ran2 & Time & Score \\

\midrule

% L-PIBT+BD & \makecell{14.76\\(0.38)} & \makecell{8.69\\(0.33)} & \makecell{7.76\\(0.17)} & \makecell{7.16\\(0.62)} & \makecell{12.73\\(0.11)} & \makecell{10.76\\(0.12)} & \makecell{0.007} & 0.89 & \makecell{42.91\\(0.07)} & \makecell{27.34\\(1.31)} & \makecell{18.84\\(0.20)} & \makecell{19.16\\(0.33)} & \makecell{53.74\\(0.10)} & \makecell{44.02\\(0.90)} & \makecell{0.024} & 0.80 \\
% \midrule
% \midrule
% PIBT+SG & \makecell{13.66\\(0.22)} & \makecell{9.91\\(0.24)} & \makecell{6.18\\(0.19)} & \makecell{4.50\\(0.74)} & \makecell{11.66\\(0.12)} & \makecell{7.46\\(0.37)} & \makecell{0.004} & 0.74 & \makecell{42.51\\(0.10)} & \makecell{39.34\\(0.11)} & \makecell{18.11\\(0.34)} & \makecell{17.62\\(0.26)} & \makecell{46.89\\(0.14)} & \makecell{25.57\\(0.51)} & \makecell{0.014} & 0.75 \\
% \midrule
% L-PIBT+SG & \makecell{16.52\\(0.09)} & \makecell{14.28\\(0.12)} & \makecell{8.85\\(0.11)} & \makecell{7.19\\(0.14)} & \makecell{12.34\\(0.11)} & \makecell{10.09\\(0.10)} & \makecell{0.007} & 0.98 & \makecell{43.98\\(0.07)} & \makecell{42.24\\(0.05)} & \makecell{22.54\\(0.17)} & \makecell{21.59\\(0.23)} & \makecell{52.49\\(0.10)} & \makecell{32.44\\(0.31)} & \makecell{0.023} & 0.85 \\
% \midrule
% \midrule

% \midrule
% \midrule
SILLM & \makecell{16.52\\(0.09)} & \makecell{\textbf{14.28}\\(0.12)} & \makecell{8.85\\(0.11)} & \makecell{\textbf{7.19}\\(0.14)} & \makecell{12.73\\(0.11)} & \makecell{\textbf{10.76}\\(0.12)} & \makecell{0.007} & 1.00 & \makecell{43.98\\(0.07)} & \makecell{42.24\\(0.05)} & \makecell{\textbf{31.00}\\(0.89)} & \makecell{\textbf{30.52}\\(0.98)} & \makecell{53.74\\(0.10)} & \makecell{\textbf{44.37}\\(0.11)} & \makecell{0.349} & 0.99 \\
\midrule
\makecell{SILLM\\(DG only)} & \makecell{14.41\\(0.16)} & \makecell{11.67\\(0.54)} & \makecell{8.34\\(0.07)} & \makecell{6.77\\(0.28)} & \makecell{11.14\\(0.13)} & \makecell{9.18\\(0.14)} & \makecell{0.029} & 0.88 & \makecell{39.38\\(0.09)} & \makecell{35.39\\(0.26)} & \makecell{\textbf{31.00}\\(0.89)} & \makecell{\textbf{30.52}\\(0.98)} & \makecell{50.48\\(0.14)} & \makecell{\textbf{44.37}\\(0.11)} & \makecell{0.721} & 0.94 \\
\midrule
WPPL \cite{Jiang2024Competition} & \makecell{\textbf{16.74}\\(0.06)} & \makecell{13.90\\(0.16)} & \makecell{\textbf{8.91}\\(0.07)} & \makecell{7.10\\(0.20)} & \makecell{\textbf{13.00}\\(0.09)} & \makecell{10.58\\(0.15)} & \makecell{1.546} & 0.99 & \makecell{\textbf{44.28}\\(0.07)} & \makecell{\textbf{42.83}\\(0.04)} & \makecell{23.74\\(0.20)} & \makecell{20.35\\(0.29)} & \makecell{\textbf{54.38}\\(0.09)} & \makecell{37.30\\(0.43)} & \makecell{4.143} & 0.88 \\
\midrule
\makecell{TrafficFlow \cite{chen2024traffic}} & \makecell{11.78\\(0.20)} & \makecell{9.10\\(0.42)} & \makecell{7.44\\(0.24)} & \makecell{5.70\\(0.43)} & \makecell{10.35\\(0.37)} & \makecell{8.48\\(0.10)} & \makecell{0.016} & 0.76 & \makecell{36.89\\(0.22)} & \makecell{27.78\\(0.88)} & \makecell{27.51\\(0.69)} & \makecell{27.03\\(0.54)} & \makecell{45.62\\(0.84)} & \makecell{33.64\\(2.32)} & \makecell{0.570} & 0.81 \\
\midrule
\makecell{PIBT \cite{okumura2022priority}} & \makecell{7.79\\(0.36)} & \makecell{4.62\\(0.10)} & \makecell{5.59\\(0.15)} & \makecell{5.12\\(0.35)} & \makecell{10.84\\(0.08)} & \makecell{6.80\\(0.48)} & \makecell{0.004} & 0.60 & \makecell{32.44\\(0.10)} & \makecell{19.39\\(2.04)} & \makecell{15.43\\(0.34)} & \makecell{15.09\\(0.26)} & \makecell{46.51\\(0.08)} & \makecell{29.08\\(1.26)} & \makecell{0.014} & 0.61 \\
\midrule
Follower \cite{skrynnik2024learn} & \makecell{10.71\\(0.10)} & \makecell{6.68\\(0.31)} & \makecell{7.21\\(0.10)} & \makecell{5.60\\(0.19)} & \makecell{10.00\\(0.07)} & \makecell{8.14\\(0.10)} & \makecell{0.051} & 0.70 & \makecell{33.79\\(0.06)} & \makecell{16.26\\(0.17)} & \makecell{7.32\\(0.32)} & \makecell{9.11\\(0.29)} & \makecell{41.39\\(0.10)} & \makecell{30.03\\(0.82)} & \makecell{4.125} & 0.52 \\
\midrule
RHCR \cite{li2021lifelong} & \makecell{10.56\\(0.27)} & \makecell{3.38\\(0.24)} & \makecell{6.27\\(0.08)} & \makecell{5.19\\(0.13)} & \makecell{12.14\\(0.09)} & \makecell{7.96\\(0.12)} & \makecell{4.829} & 0.66 & - & - & - & - & - & - & - & - \\
\midrule
SCRIMP \cite{wang2023scrimp} & \makecell{1.01\\(0.10)} & \makecell{0.75\\(0.06)} & \makecell{1.42\\(0.18)} & \makecell{0.53\\(0.05)} & \makecell{6.29\\(0.55)} & \makecell{2.08\\(0.27)} & \makecell{1.982} & 0.17 & - & - & - & - & - & - & - & - \\
\bottomrule
% \bottomrule
    \end{tabular}
    }
\vspace{-0.2cm}
\end{table*}

\section{Experiments}
\label{ss: experiments}

Applying imitation learning directly in large-scale settings is possible, but complex engineering is required to deal with the large memory consumption during training. Therefore, we first downscaled the large maps to small ones but kept the original obstacle patterns. We trained a neural policy with $600$ agents on each small map ($500$ timesteps for all maps) but evaluated it with $10,000$ agents on the corresponding large map ($3200$ timesteps for Sortation and Warehouse and $2500$ timesteps for others). 
% We simulate $500$ steps on all small maps, and $2500$ steps on all large maps except the Sortation and Warehouse where we simulate $3200$ steps so that the number of steps is roughly $4$ times the sum of a map' height and width.
% The agent densities are $16.3\% \sim 47.0\%$ on small maps and $17.0\% \sim 35.5\%$ on large maps.
More details of each map are covered in \Cref{tab:instance}. We provide in the project webpage\footnote{https://diligentpanda.github.io/SILLM/} our code and appendix, including the evaluation of different agent numbers and the benchmark used by Follower \cite{skrynnik2024learn}.

% More experiments and settings are covered in the Appendix, including the evaluation of different agent numbers and another benchmark used by Follower \cite{skrynnik2024learn}.

\subsection{Main Results}
\label{ss: main results}
This subsection compares different variants of our solver, SILLM, with other state-of-the-art search- and learning-based solvers. We use SCRIMP \cite{wang2023scrimp} and Follower \cite{skrynnik2024learn} as learning-based baselines and PIBT \cite{okumura2022priority}, RHCR \cite{li2021lifelong}, and TrafficFlow \cite{chen2024traffic} as search-based baselines. In addition, we also compare SILLM with WPPL \cite{Jiang2024Competition}, the winning solution of the League of Robot Runners competition \cite{chan2024league}. To compare different solvers more easily, we compute scores in the same way as the competition: for each instance, we collect the best throughput achieved among all the solvers evaluated in this paper. Then, a score between $0$ and $1$ is computed as the throughput of the solver divided by the best throughput.
% \begin{align*}
%     \text{score}=\frac{\text{The throughput of the solver}}{\text{The best throughput}}
% \end{align*}
% The solvers are then ranked by their average scores over all instances.

\begin{figure}[tbp]
  \centering
  \begin{minipage}{0.5\textwidth}
    \centering
        \includegraphics[width=0.7\linewidth]{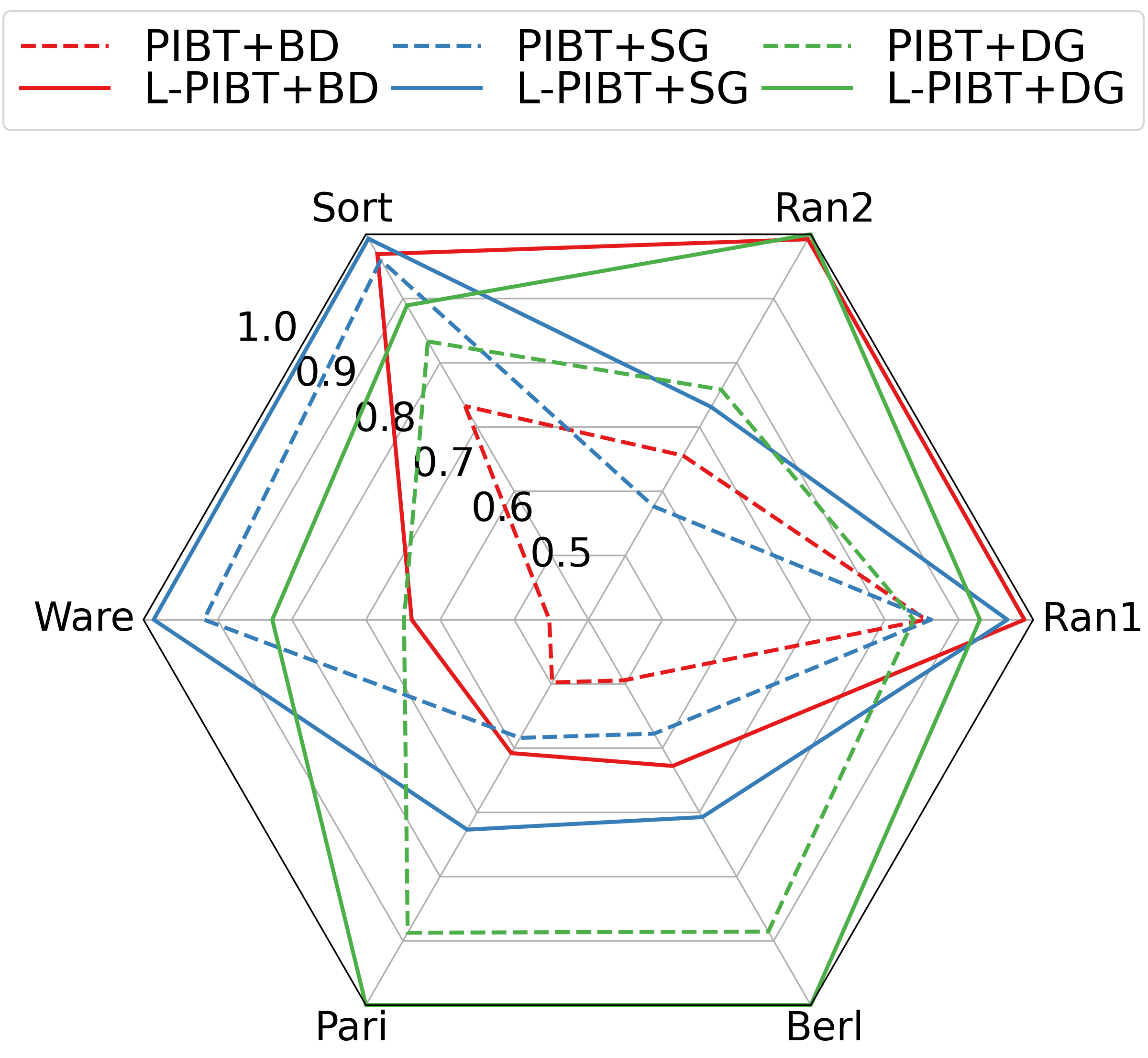}
        \label{fig:enter-label}
  \end{minipage}%
  
    \vspace{0.3cm}
    
  \begin{minipage}{0.5\textwidth}
    \centering
    \resizebox{0.95\textwidth}{!}{
    \begin{tabular}{c|c|c|c}
      \toprule
      Score (Time) & BD & SG & DG \\
      \midrule
      PIBT   & 0.61 (0.014s) & 0.75 (0.014s) & 0.81 (0.570s)\\
      L-PIBT & 0.80 (0.024s) & 0.85 (0.023s) & 0.94 (0.721s)\\
      \bottomrule
    \end{tabular}
    % \captionof{table}{Sample table}
    }
  \end{minipage}
  \caption{Comparison of Learnable-PIBT (L-PIBT) and PIBT with different global guidance on large instances. The radar plot shows the score for each instance. The table reports the average score and the average planning time per timestep.}
  \label{fig: radar}
   % \vspace{-0.3cm}
\end{figure}

First, we compare the performance of PIBT and Learnable PIBT with different global guidance in \Cref{fig: radar}. Learnable PIBT (solid line) consistently achieves better throughput than its counterpart (dashed line). The increase in the planning time is not negligible but acceptable. Different types of global guidance excel at different kinds of maps in our experiments. For example, Static Guidance, implemented as a criss-cross highway, fits the criss-cross patterns of sortation and warehouse maps better. For random instances, normal Backward Dijkstra actually performs better. The potential reasons are: (1) uniformly generated random structures implicitly make agents select more distributed paths and (2) with lower agent density and potentially less congestion, normal Backward Dijkstra is a more accurate estimation of the future steps. 
%Regarding the average score on large instances, Dynamic Guidance is the best, though it takes much longer to plan the guide paths at each step, while the other two could precompute the heuristic values.
Dynamic Guidance performs well across all maps and achieves the best results on city-type maps (i.e., Paris and Berlin), where obstacle patterns are less uniform.

Thus, we select the best performance achieved by Learnable PIBT with one of the global guidance to form the results of our solver, SILLM, in \Cref{tab:main}. We also report the performance of using Dynamic Guidance only as SILLM (DG only). Then, we compare SILLM with other search- and learning-based algorithms in \Cref{tab:main}. Regarding the score, SILLM outperforms all other learning-based and search-based methods. Specifically, on large instances, SILLM almost doubles the score of the previous state-of-the-art learning-based method, Follower \cite{skrynnik2024learn}, and also significantly outperforms the state-of-the-art search-based method,  TrafficFlow \cite{chen2024traffic}.
Compared to WPPL \cite{Jiang2024Competition}, SILLM performs either closely or much better on all maps and is much faster. With Dynamic Guidance only, SILLM (DG only) also achieves a better score than any other baselines on large instances.
%The main reason is that SILLM incorporates dynamic global guidance. 
% Notably, SILLM is also much faster than WPPL. 
%The main reason our SILLM outperforms WPPL is the incorporation of Dynamic Guidance. We also try to directly combine WPPL and Dynamic Guidance. But without acceleration from neural policy, its speed is too slow (more than 10 seconds per step). 
Earlier methods like RHCR \cite{li2021lifelong} and SCRIMP \cite{wang2023scrimp} cannot scale to the large scale because of their heavy computation. 
%RHCR and SCRIMP take minutes per step on large instances and are too costly to evaluate, failing to meet real-time requirements. 
RHCR is slow due to its use of a bounded suboptimal but relatively slow search method, ECBS \cite{barer2014suboptimal}, for each planning window. %Its performance on the small warehouse map is poor because of frequent timeouts exceeding the 10-second limit. % SCRIMP is slow because its single-step collision avoidance based on action resampling is very time-consuming when agents' movements are mutually dependent.
SCRIMP's time consumption stems primarily from its single-step collision avoidance strategy, which requires action resampling and team state value calculation. 
%In large-scale tasks involving 10,000 agents, it takes around 10 minutes for each step due to the excessive collisions that need to be resolved.

% % \subsection{Generalization }
\subsection{Ablation Study}
\label{ss: ablation study}
% \subsubsection{Communication Module}
% \subsubsection{Reinforcement Learning}

\begin{figure}
    \centering
    \includegraphics[width=1.0\linewidth]{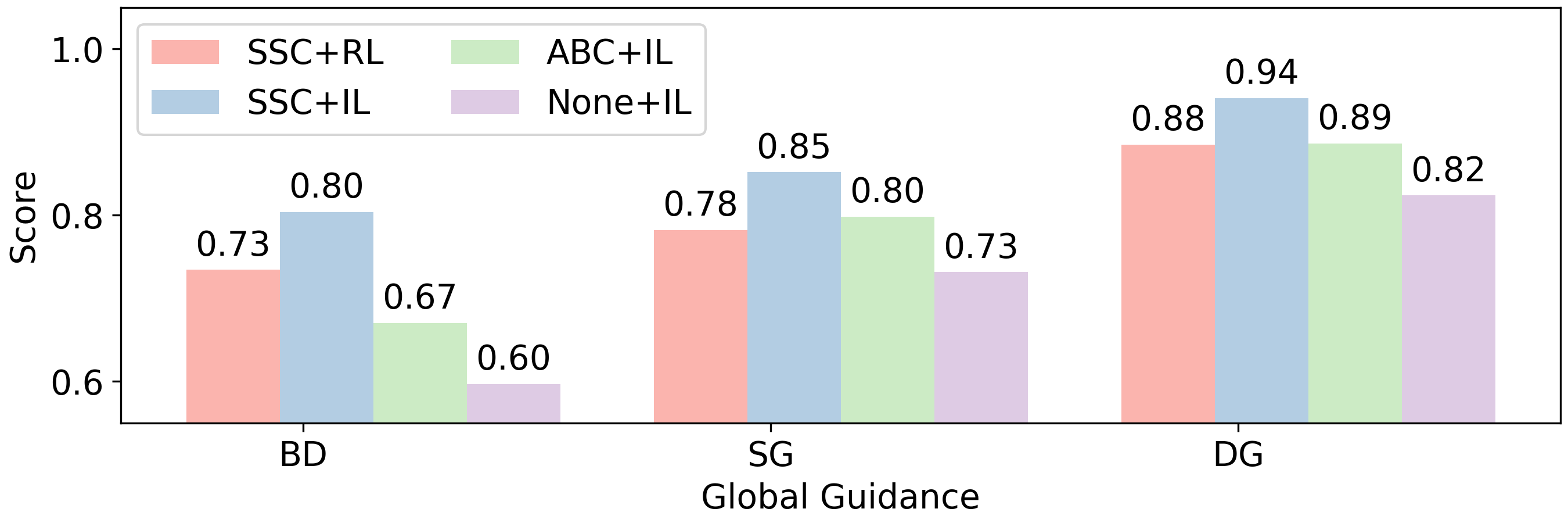}
    \caption{Ablation studies on large maps.}
    \label{fig:ablation_comm}
\vspace{-0.3cm}
\end{figure}

\Cref{fig:ablation_comm} compares our spatially sensitive communication (SSC) with attention-based communication (ABC) and no communication (None) trained by imitation learning (IL). SSC consistently outperforms ABC and None, indicating the importance of precise spatial reasoning in LMAPF. 
%Notably, without communication, learning almost doesn't improve the performance in our case, which emphasizes the importance of communication. 
We further compare our IL with a reinforcement learning (RL) implementation based on MAPPO \cite{yu2022surprising} and find that our IL outperforms the simple MAPPO. We also tried to apply MAPPO after IL but did not notice any improvement. We believe more sophisticated RL methods are needed.
%, especially those can properly distribute team-level rewards to individuals, which is left for future work.

% \begin{figure}
%     \centering
%     \includegraphics[width=0.8\linewidth]{figures//exp//ablation/ablation_score_communication_large.pdf}
%     \caption{Ablation studies on communication modules.}
%     \label{fig:ablation_comm}
% \end{figure}

% \begin{figure}
%     \centering
%     \includegraphics[width=0.8\linewidth]{figures//exp//ablation/ablation_score_RL_large.pdf}
%     \caption{Ablation study on training methods.}
%     \label{fig:ablation_training}
% \end{figure}

% \begin{figure}
%  \begin{subfigure}[b]{0.5\textwidth}
%       \centering
%       \includegraphics[width=0.8\textwidth]{figures//exp//ablation/ablation_score_communication_large.pdf}
%       \caption{On Communication modules.}
%       \label{fig:ablation_comm}
%     \end{subfigure}%
    
%     %\par\bigskip 
%     \begin{subfigure}[b]{0.5\textwidth}
%       \centering
%       \includegraphics[width=0.8\textwidth]{figures//exp//ablation/ablation_score_RL_large.pdf}
%       \caption{On training methods.}
%       \label{fig:ablation_training}
%     \end{subfigure}%

% \caption{Ablation studies.}

% \end{figure}

% \subsection{Validation on Learn-to-Follow Benchmark}
\subsection{Real-World Mini Example}
Due to hardware and software limitations, we validate our solver with $10$ real robots and $100$ virtual robots in challenging mini warehouse environments with multiple corridors. Details can be found on the project webpage.

\section{Conclusion}
% In this work, we investigate how to scale learning-based solvers to manage a large number of agents. Specifically, we studies the objective of imitation learning, the design of communication module, different types of global guidance encoded in the local observation and the efficient single-step collision resolution. As a result, our proposed learning-based solver, SILLM, could manage as many as 10,000 agents in various maps. It outperforms other state-of-art learning- and search-based
% solvers and validates the potential of applying learning for large-scale LMAPF instances.
In this work, we show how to scale learning-based solvers to manage a large number of agents within a short planning time. Specifically, we design a unique communication module, incorporate efficient single-step collision resolution and different types of global guidance, and apply scalable imitation learning from a scalable search-based solver. As a result, our proposed solver, SILLM, can effectively plan paths for 10,000 agents at each timestep in less than 1 second for various maps. It outperforms previously best learning- and search-based solvers and validates the potential of applying learning for large-scale LMAPF instances.
Future work will explore how to further improve SILLM using RL.
% % future work:

\section*{Acknowledgments}
The research was supported by the National Science Foundation (NSF) under grant number \#2328671, a gift from Amazon, and an Amazon research award.
% The views and conclusions contained in this document are those of the authors and should not be interpreted as representing the official policies, either expressed or implied, of the sponsoring organizations, agencies, or the U.S. government.

%                                   % the textheight of the last page by a suitable amount.
%                                   % This command does not take effect until the next page
%                                   % so it should come on the page before the last. Make
%                                   % sure that you do not shorten the textheight too much.

%%%%%%%%%%%%%%%%%%%%%%%%%%%%%%%%%%%%%%%%%%%%%%%%%%%%%%%%%%%%%%%%%%%%%%%%%%%%%%%%

%%%%%%%%%%%%%%%%%%%%%%%%%%%%%%%%%%%%%%%%%%%%%%%%%%%%%%%%%%%%%%%%%%%%%%%%%%%%%%%%

%%%%%%%%%%%%%%%%%%%%%%%%%%%%%%%%%%%%%%%%%%%%%%%%%%%%%%%%%%%%%%%%%%%%%%%%%%%%%%%%

% \section*{ACKNOWLEDGMENT}
% ...

%%%%%%%%%%%%%%%%%%%%%%%%%%%%%%%%%%%%%%%%%%%%%%%%%%%%%%%%%%%%%%%%%%%%%%%%%%%%%%%%

% References are important to the reader; therefore, each citation must be complete and correct. If at all possible, references should be commonly available publications.

\clearpage

\bibliographystyle{IEEEtran}
\bibliography{ICRA25}

\clearpage
\section*{APPENDIX}
\label{s: appendix}

\subsection{Visualization of all maps}
\begin{figure}[tbh]
    \centering
    \begin{subfigure}[b]{0.5\textwidth}
        \setlength{\abovecaptionskip}{5pt}
        \setlength{\belowcaptionskip}{7.5pt}
      \centering
      \includegraphics[width=1\textwidth]{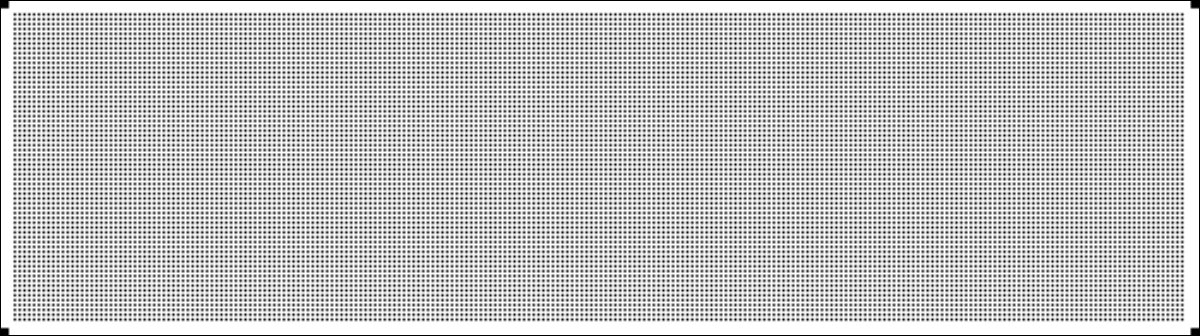}
      \caption{Sortation}
      % \label{fig:random_corner}
    \end{subfigure}%
    
    %\par\bigskip 
    \begin{subfigure}[b]{0.5\textwidth}
        \setlength{\abovecaptionskip}{5pt}
        \setlength{\belowcaptionskip}{7.5pt}
      \centering
      \includegraphics[width=1\textwidth]{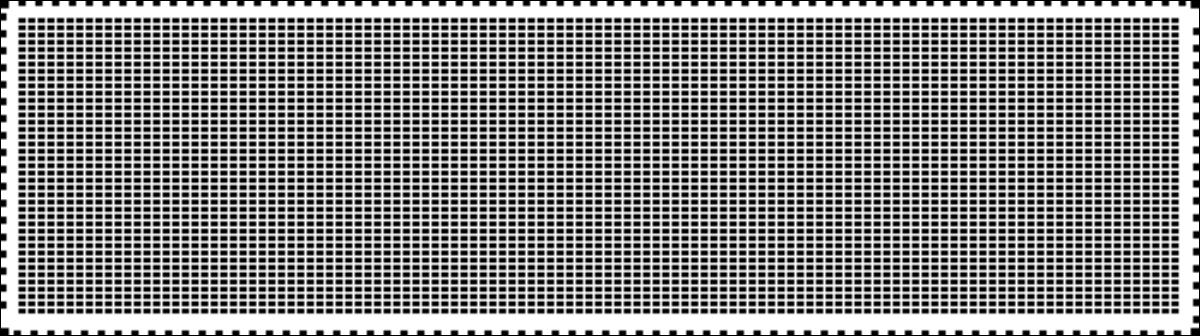}
      \caption{Warehouse}
      % \label{fig:random_corner}
    \end{subfigure}%
    
    %\par\bigskip 
    \begin{subfigure}[b]{0.24\textwidth}
            \setlength{\abovecaptionskip}{5pt}
        \setlength{\belowcaptionskip}{7.5pt}
      \centering
      \includegraphics[width=1\textwidth]{figures/maps/Paris_1_256.pdf}
      \caption{Paris}
      % \label{fig:random_corner}
    \end{subfigure}%
    \hfill
    \begin{subfigure}[b]{0.24\textwidth}
            \setlength{\abovecaptionskip}{5pt}
        \setlength{\belowcaptionskip}{7.5pt}
      \centering
      \includegraphics[width=1\textwidth]{figures/maps/Berlin_1_256.pdf}
      \caption{Berlin}
      % \label{fig:random_corner}
    \end{subfigure}%

    %\par\bigskip 
    \begin{subfigure}[b]{0.24\textwidth}
            \setlength{\abovecaptionskip}{5pt}
        \setlength{\belowcaptionskip}{7.5pt}
      \centering
      \includegraphics[width=1\textwidth]{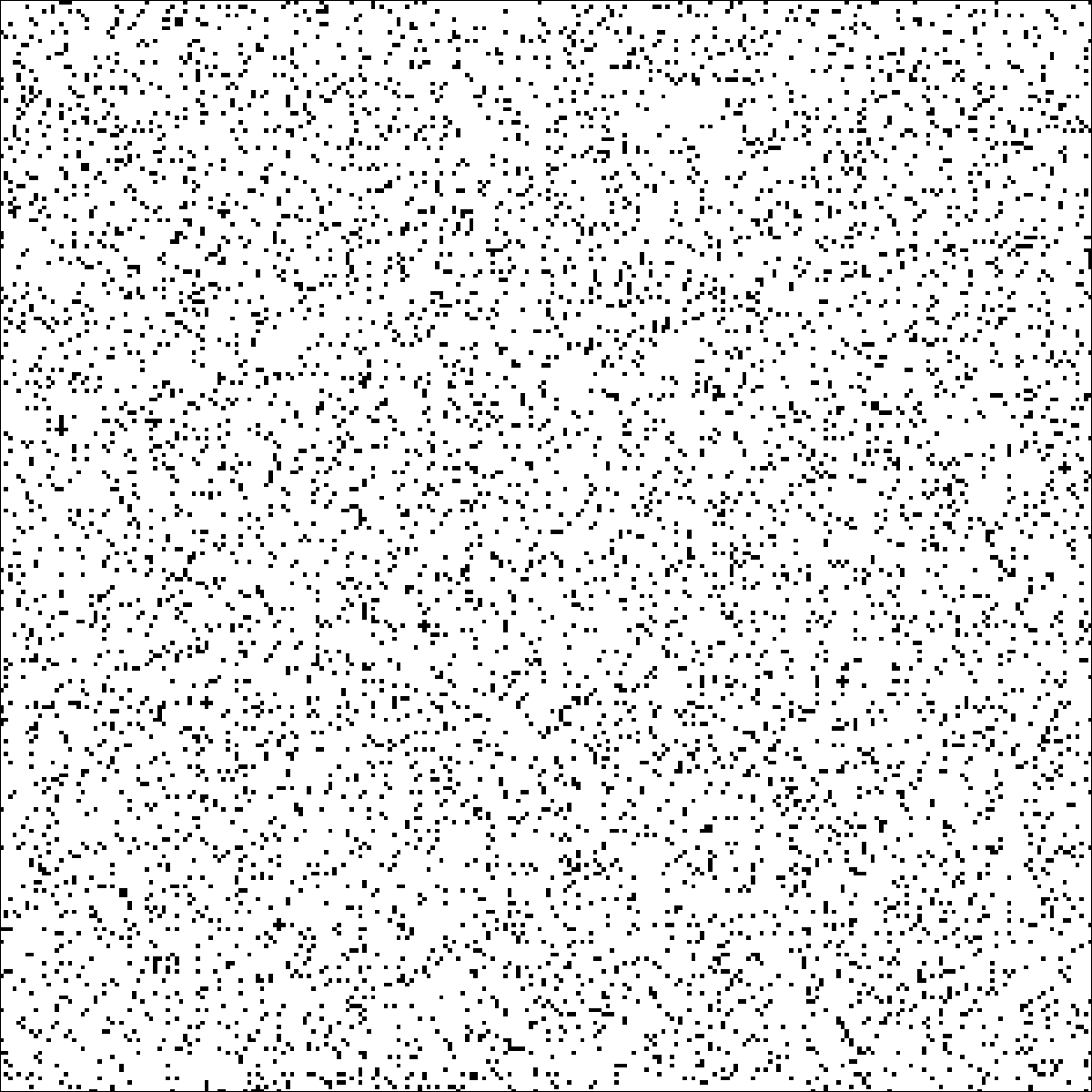}
      \caption{Random1}
      % \label{fig:random_corner}
    \end{subfigure}%
    \hfill
    \begin{subfigure}[b]{0.24\textwidth}
            \setlength{\abovecaptionskip}{5pt}
        \setlength{\belowcaptionskip}{7.5pt}
      \centering
      \includegraphics[width=1\textwidth]{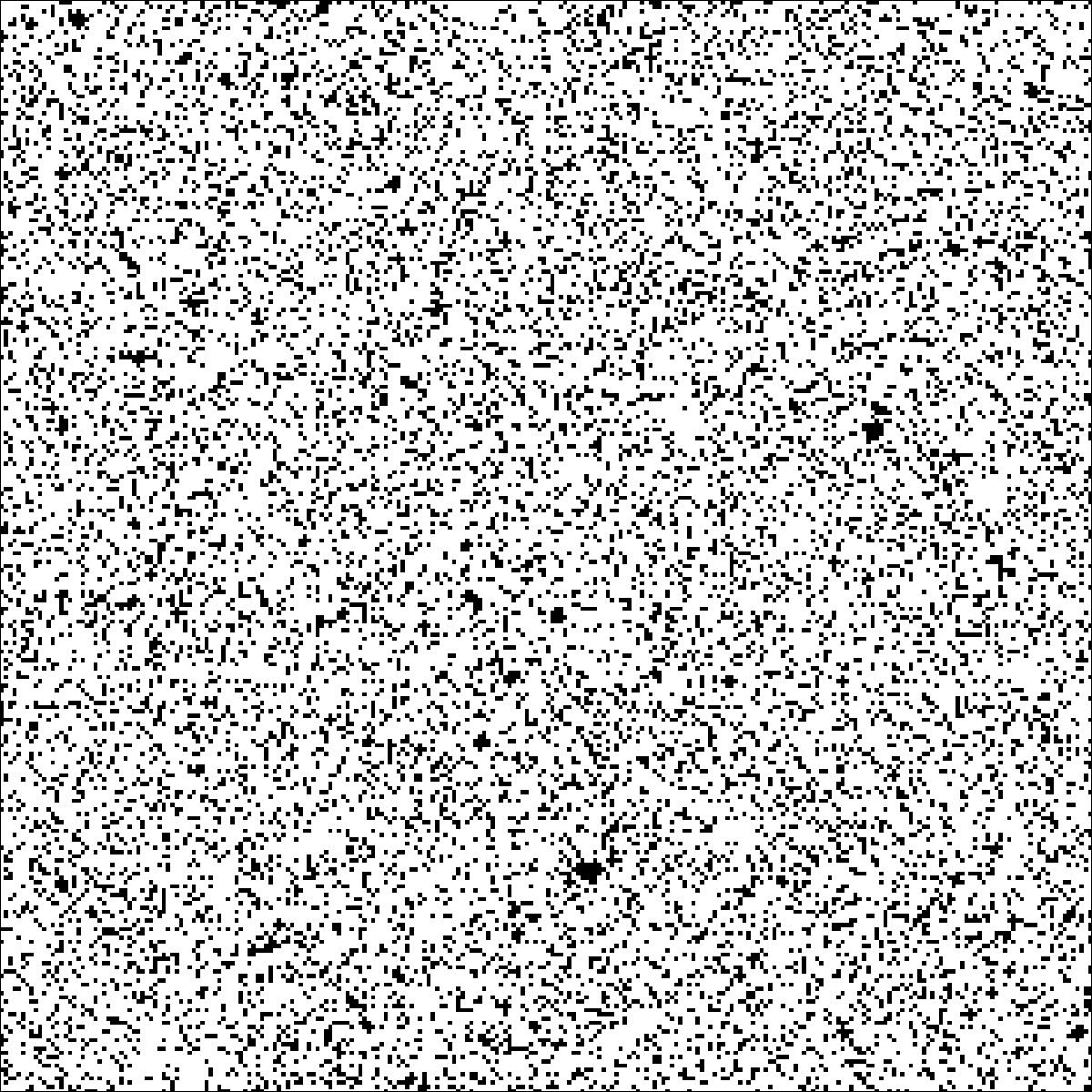}
      \caption{Random2}
      % \label{fig:random_corner}
    \end{subfigure}%
    
    \caption{Large maps with $10,000$ agents for evaluation.}
    \label{fig: large maps}
\end{figure}

\begin{figure}[tbh]
    \centering
    \begin{subfigure}[b]{0.24\textwidth}
            \setlength{\abovecaptionskip}{5pt}
        \setlength{\belowcaptionskip}{7.5pt}
      \centering
      \includegraphics[width=1\textwidth]{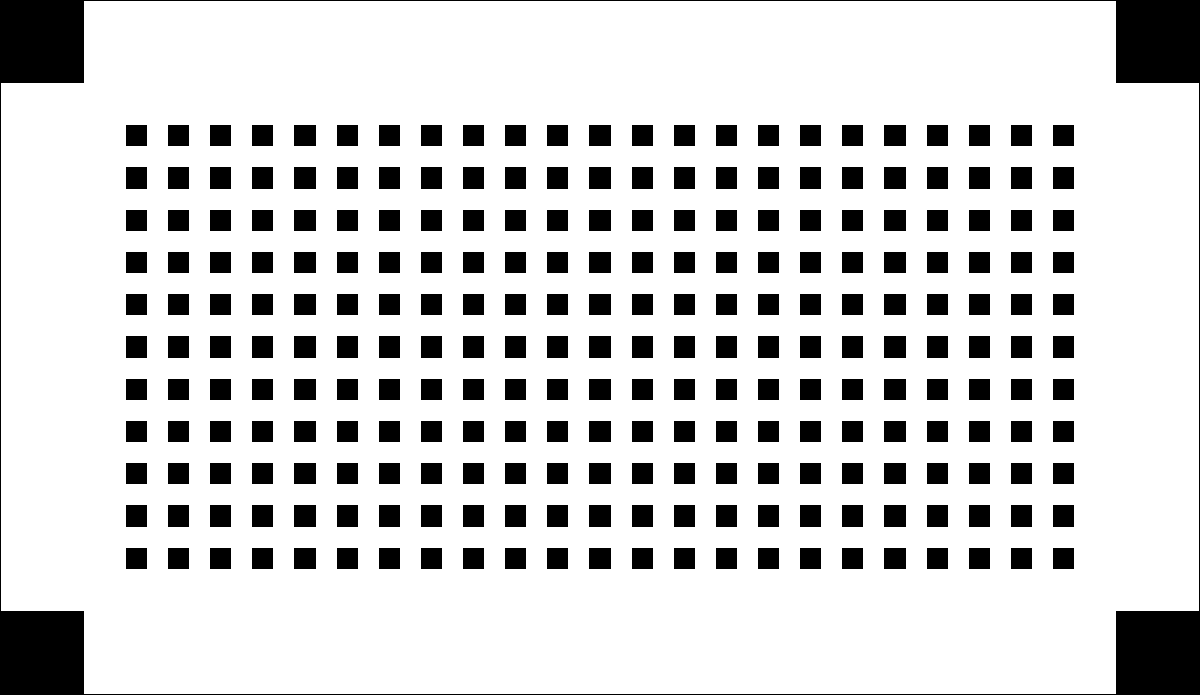}
      \caption{Sortation}
      % \label{fig:random_corner}
    \end{subfigure}%
    \hfill
    %\par\bigskip 
    \begin{subfigure}[b]{0.24\textwidth}
            \setlength{\abovecaptionskip}{5pt}
        \setlength{\belowcaptionskip}{7.5pt}
      \centering
      \includegraphics[width=1\textwidth]{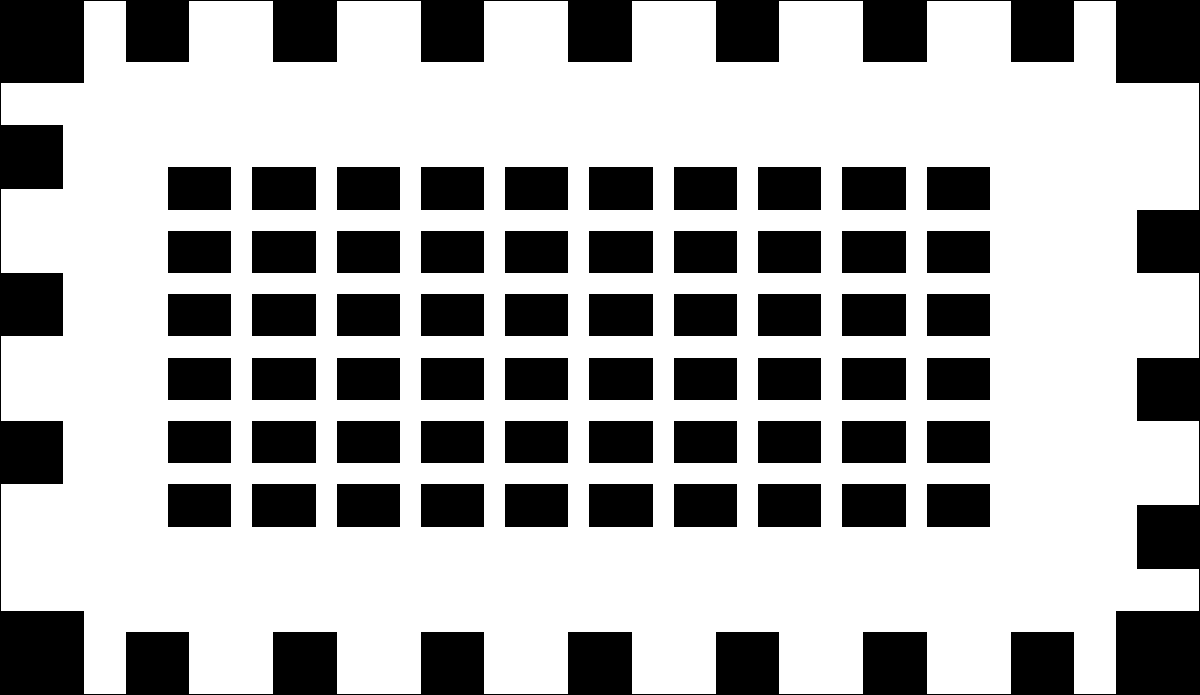}
      \caption{Warehouse}
      % \label{fig:random_corner}
    \end{subfigure}%
    \hfill    %\par\bigskip 
    
    %\par\bigskip 
    \begin{subfigure}[b]{0.24\textwidth}
            \setlength{\abovecaptionskip}{5pt}
        \setlength{\belowcaptionskip}{7.5pt}
      \centering
      \includegraphics[width=1\textwidth]{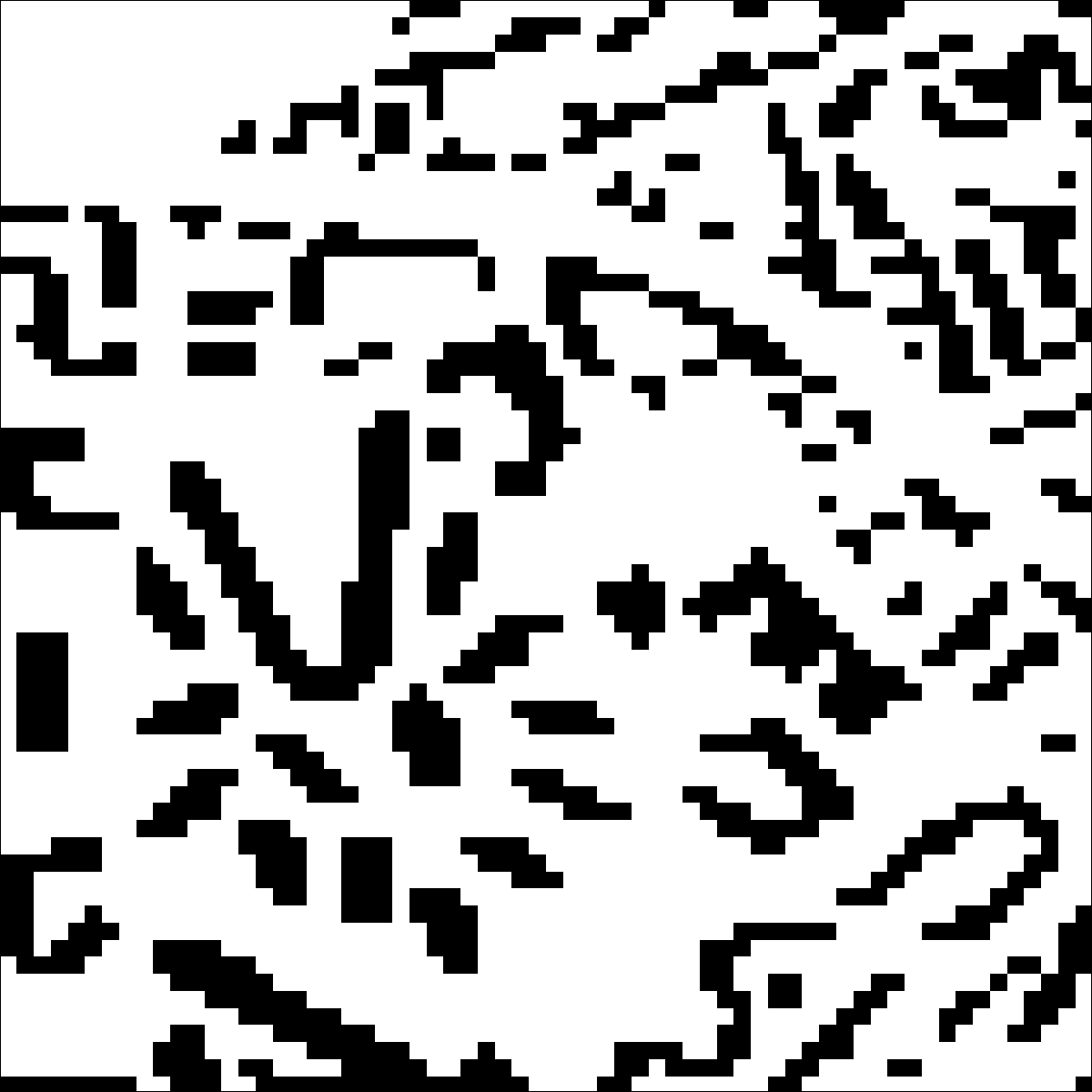}
      \caption{Paris}
      % \label{fig:random_corner}
    \end{subfigure}%
    \hfill
    \begin{subfigure}[b]{0.24\textwidth}
            \setlength{\abovecaptionskip}{5pt}
        \setlength{\belowcaptionskip}{7.5pt}
      \centering
      \includegraphics[width=1\textwidth]{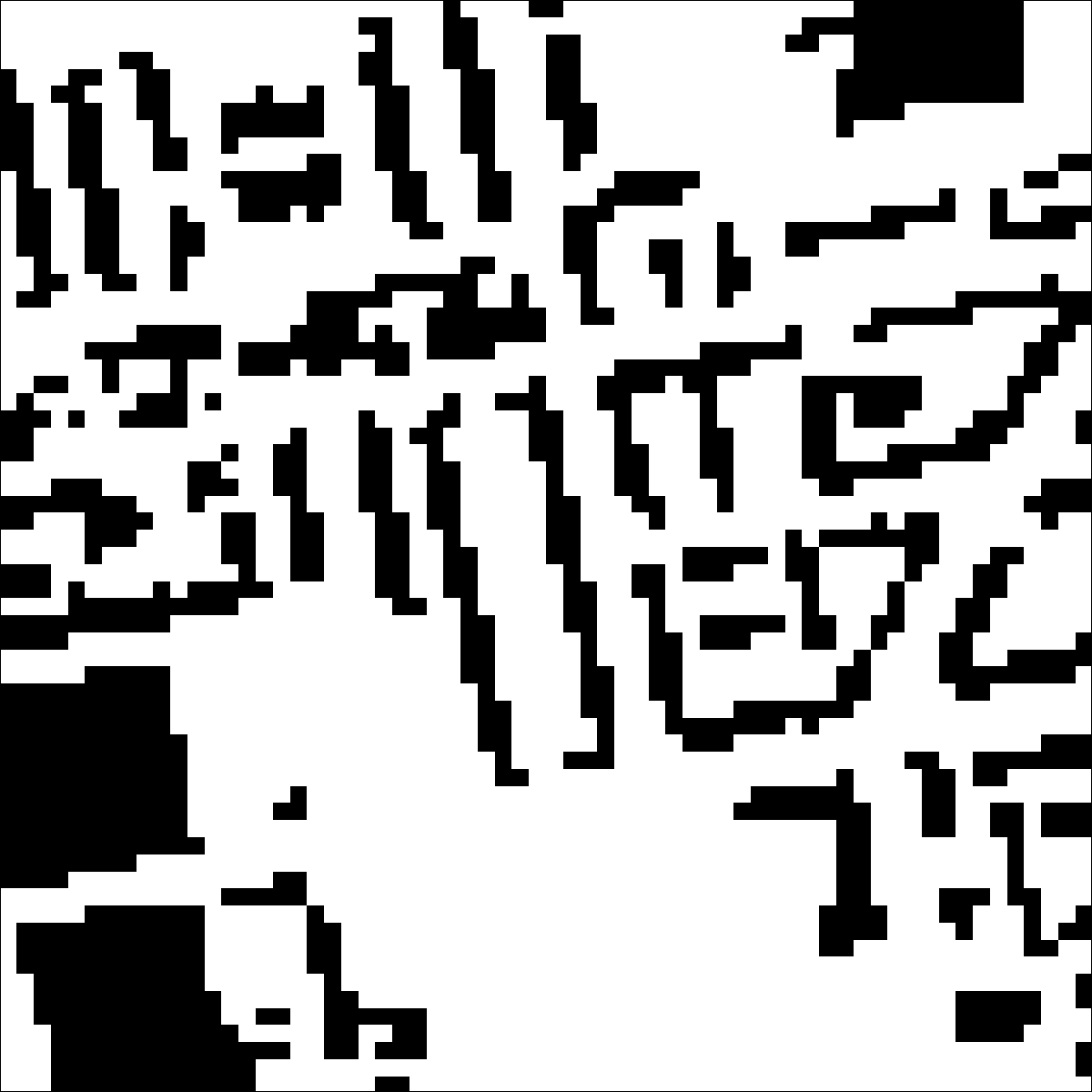}
      \caption{Berlin}
      % \label{fig:random_corner}
    \end{subfigure}%

    %\par\bigskip 
    \begin{subfigure}[b]{0.24\textwidth}
            \setlength{\abovecaptionskip}{5pt}
        \setlength{\belowcaptionskip}{7.5pt}
      \centering
      \includegraphics[width=1\textwidth]{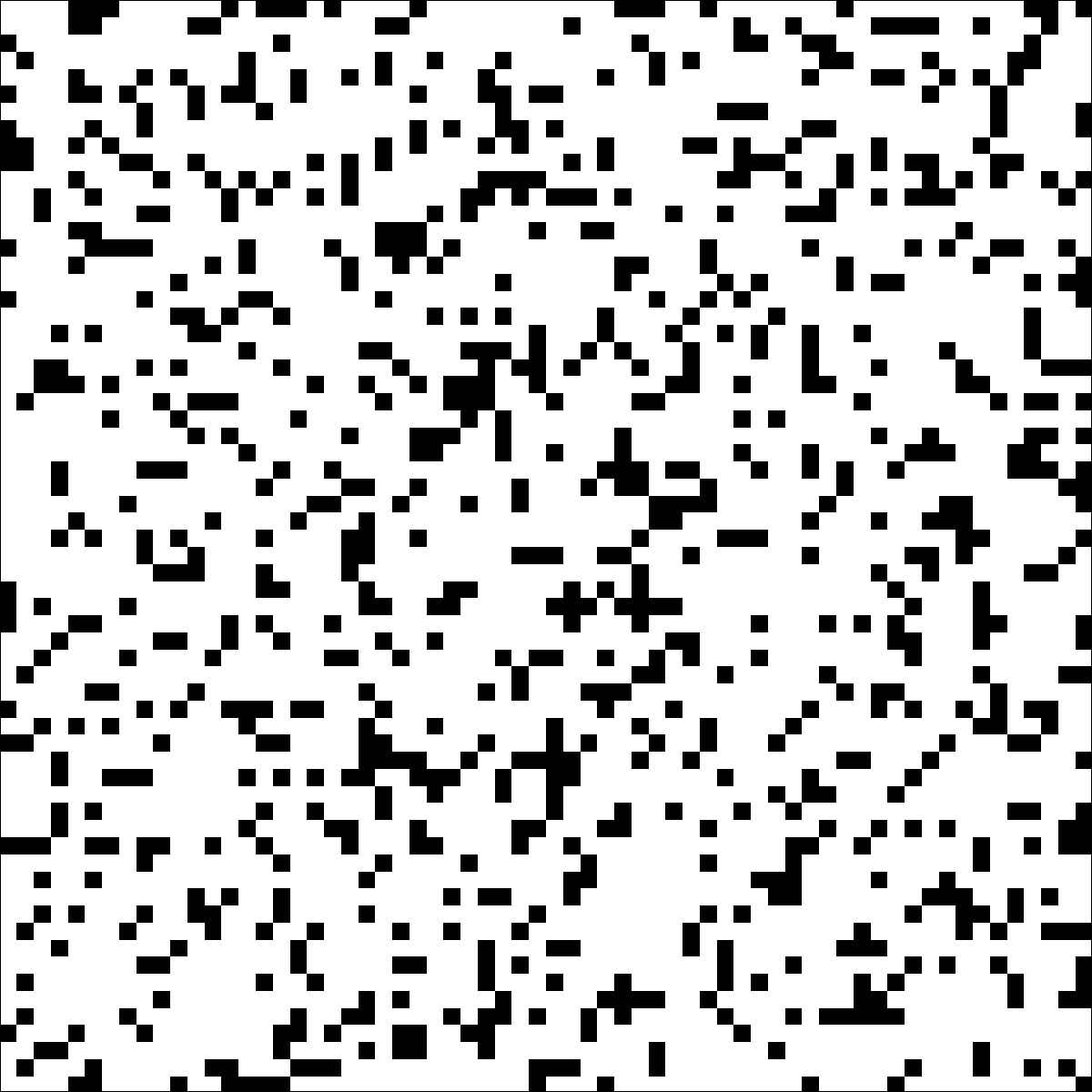}
      \caption{Random1}
      % \label{fig:random_corner}
    \end{subfigure}%
    \hfill
    \begin{subfigure}[b]{0.24\textwidth}
            \setlength{\abovecaptionskip}{5pt}
        \setlength{\belowcaptionskip}{7.5pt}
      \centering
      \includegraphics[width=1\textwidth]{figures/maps/random_256_20_small.pdf}
      \caption{Random2}
      % \label{fig:random_corner}
    \end{subfigure}%
    
    \caption{Small maps with $600$ agents for training.}
    \label{fig: small maps}
\end{figure}

We visualize all the large maps for evaluation in \Cref{fig: large maps} and all the down-scaled small maps for training in \Cref{fig: small maps}, which keep the obstacle patterns in the 
 corresponding large maps.

\subsection{Evaluation with Different Numbers of Agents}
\label{ss: eval_diff_num_agents}
\begin{figure}[!b]
    \centering
    \begin{subfigure}[b]{0.45\textwidth}
      \centering
      \includegraphics[width=1\textwidth]{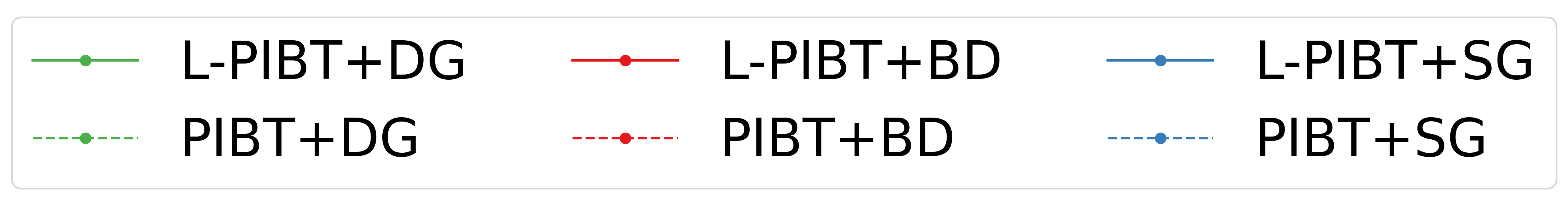}
      % \caption{Random 20*20}
      % \label{fig:random_corner}
    \end{subfigure}%

    \begin{subfigure}[b]{0.24\textwidth}
      \centering
      \includegraphics[width=1\textwidth]{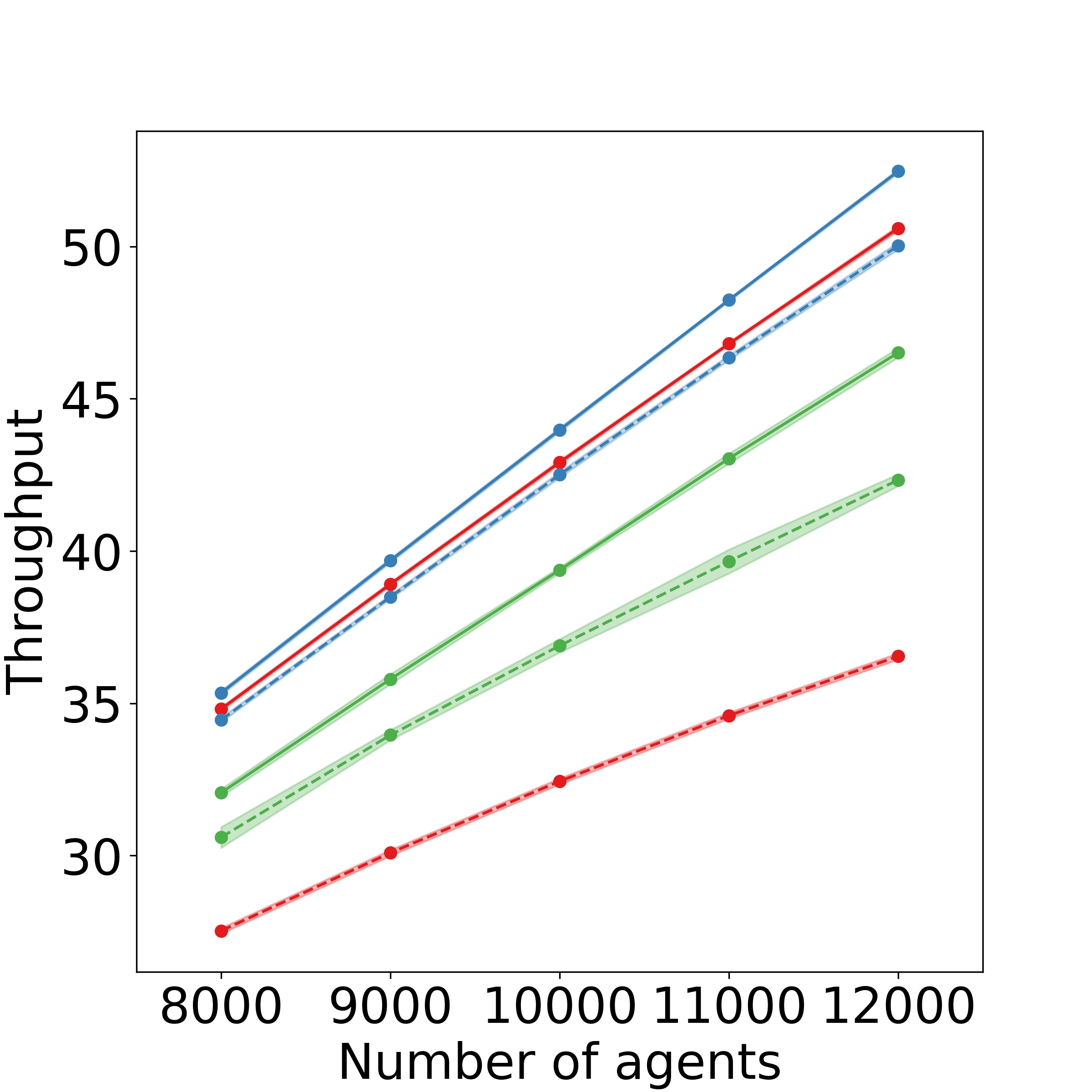}
      \caption{Sortation}
      % \label{fig:random_corner}
    \end{subfigure}%
    \hfill
    %\par\bigskip 
    \begin{subfigure}[b]{0.24\textwidth}
      \centering
      \includegraphics[width=1\textwidth]{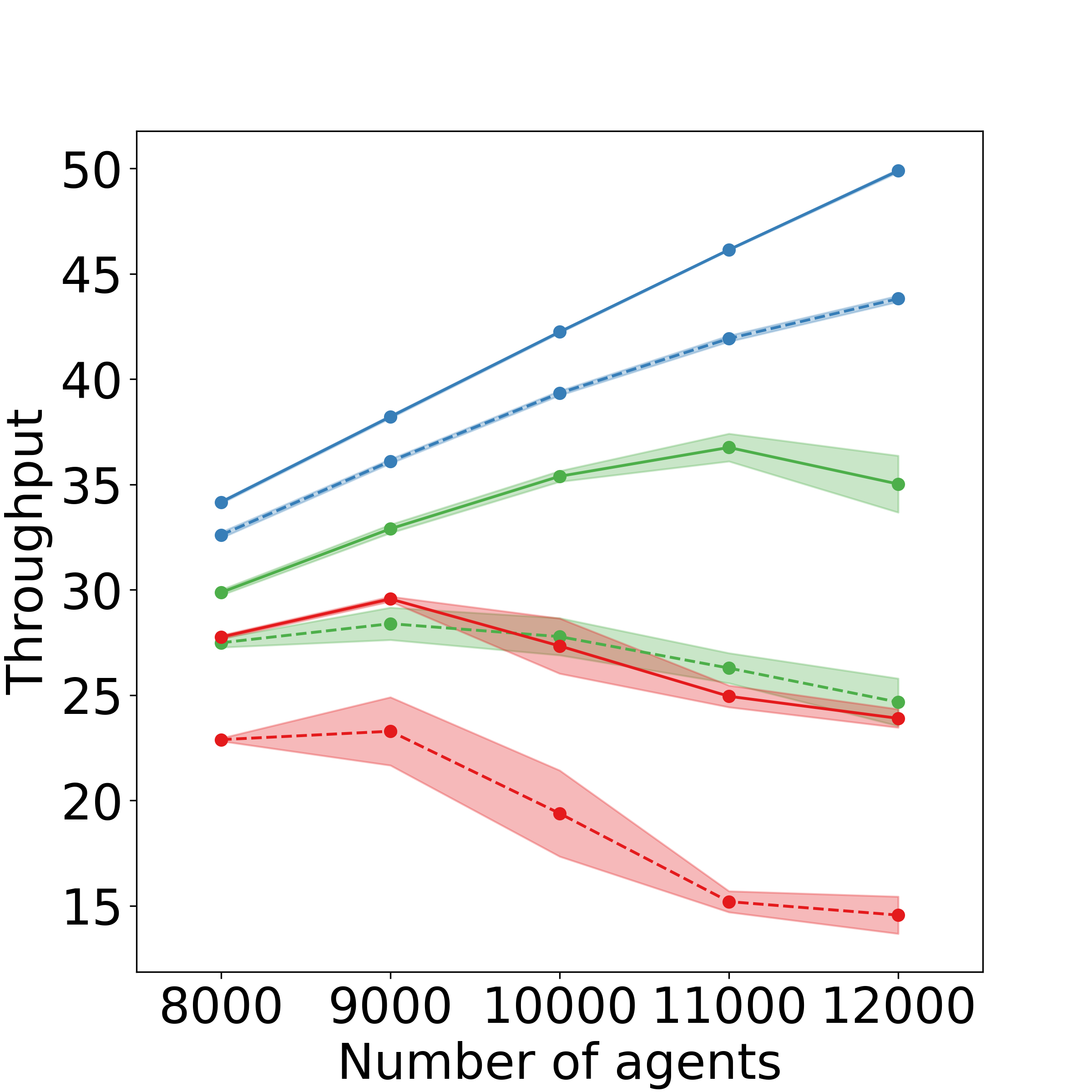}
      \caption{Warehouse}
      % \label{fig:random_corner}
    \end{subfigure}%
    \hfill    %\par\bigskip 
    
    %\par\bigskip 
    \begin{subfigure}[b]{0.24\textwidth}
      \centering
      \includegraphics[width=1\textwidth]{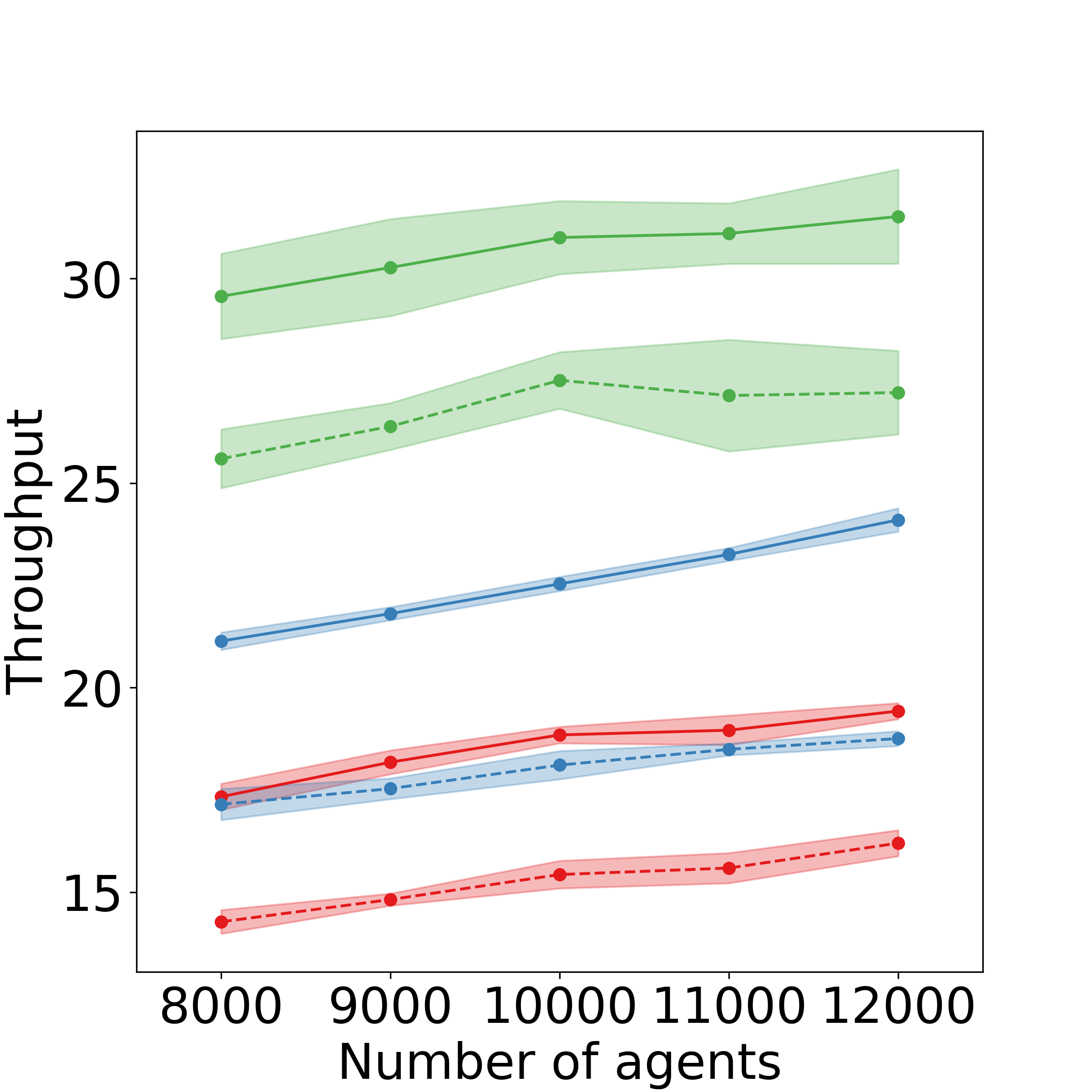}
      \caption{Paris}
      % \label{fig:random_corner}
    \end{subfigure}%
    \hfill
    \begin{subfigure}[b]{0.24\textwidth}
      \centering
      \includegraphics[width=1\textwidth]{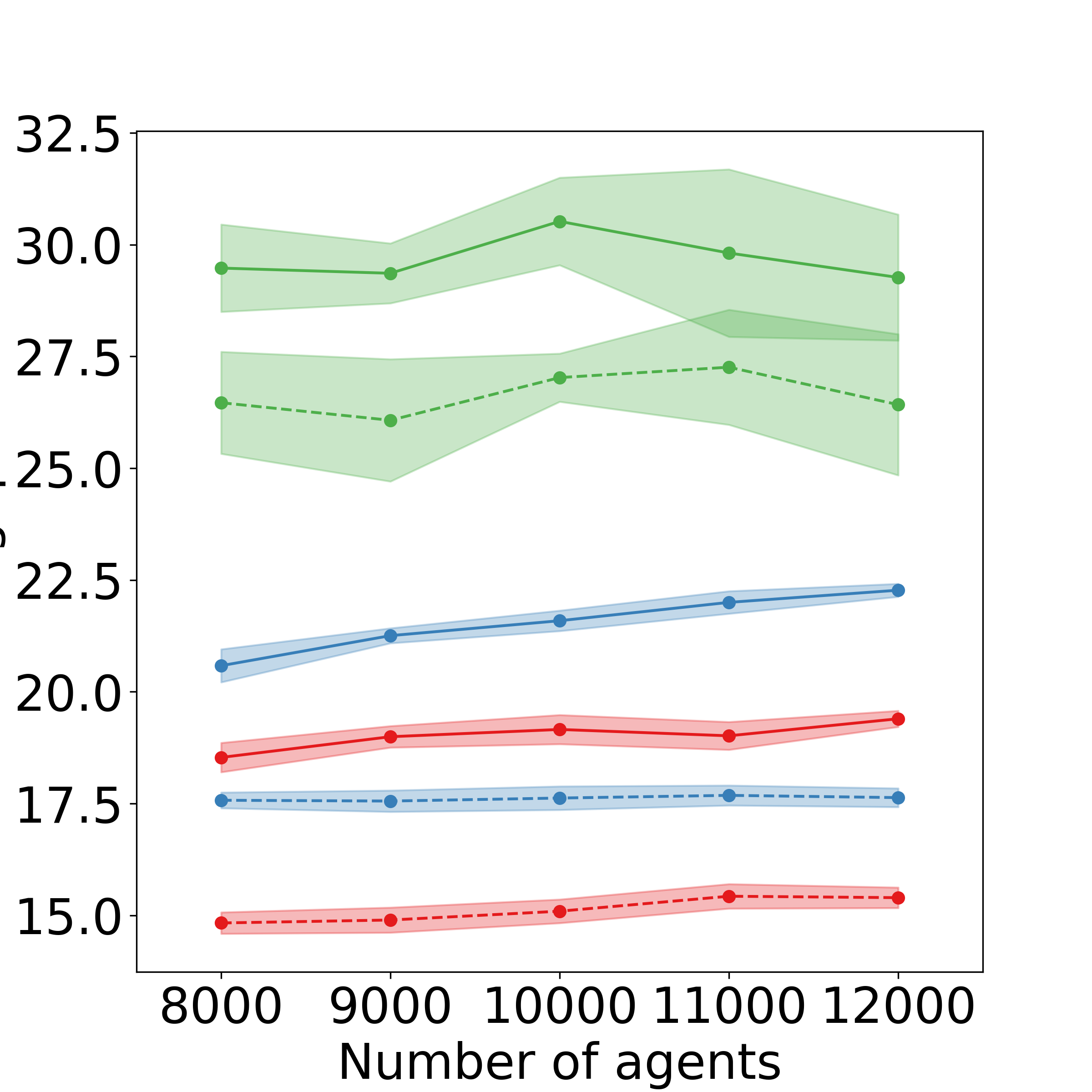}
      \caption{Berlin}
      % \label{fig:random_corner}
    \end{subfigure}%

    %\par\bigskip 
    \begin{subfigure}[b]{0.24\textwidth}
      \centering
      \includegraphics[width=1\textwidth]{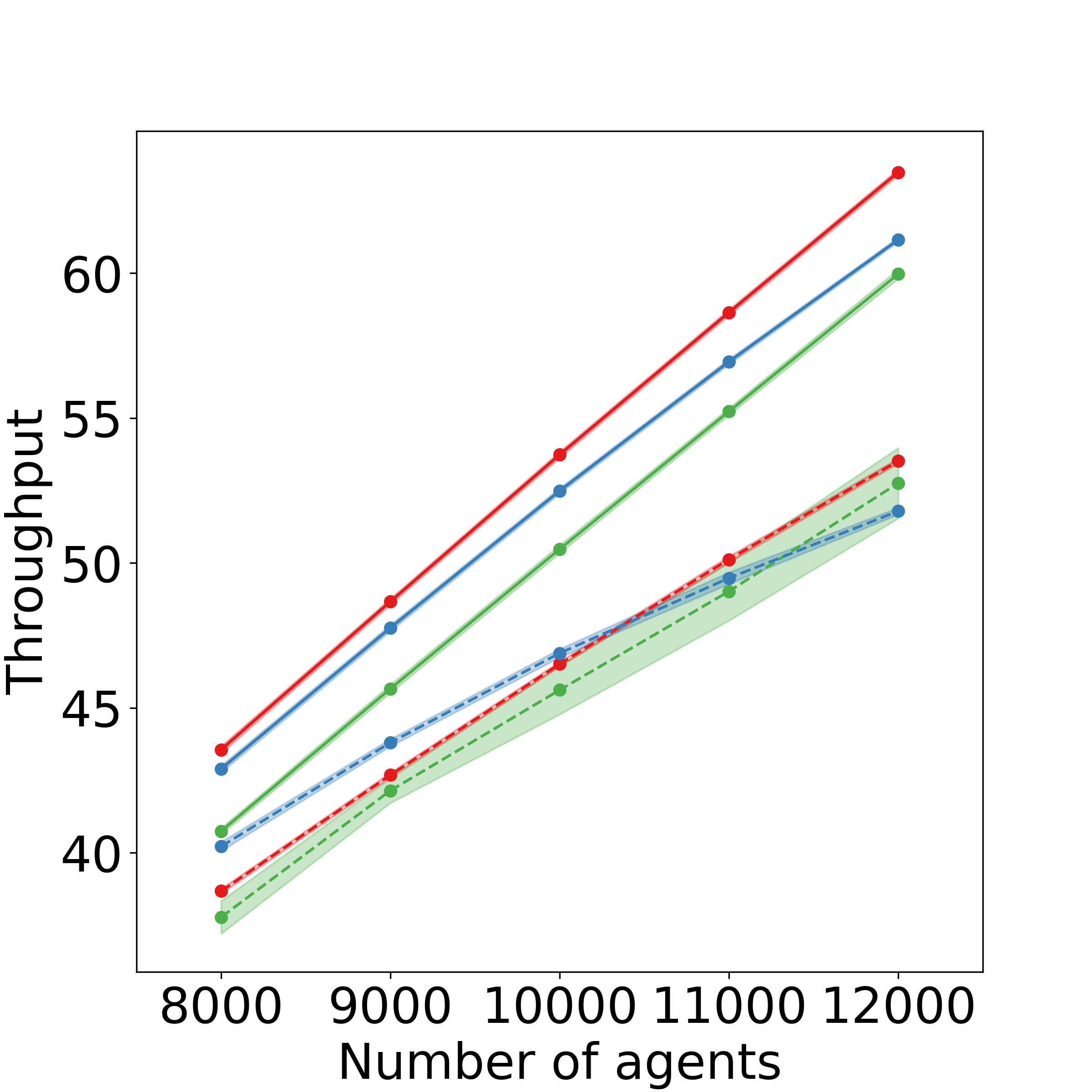}
      \caption{Random1}
      % \label{fig:random_corner}
    \end{subfigure}%
    \hfill
    \begin{subfigure}[b]{0.24\textwidth}
      \centering
      \includegraphics[width=1\textwidth]{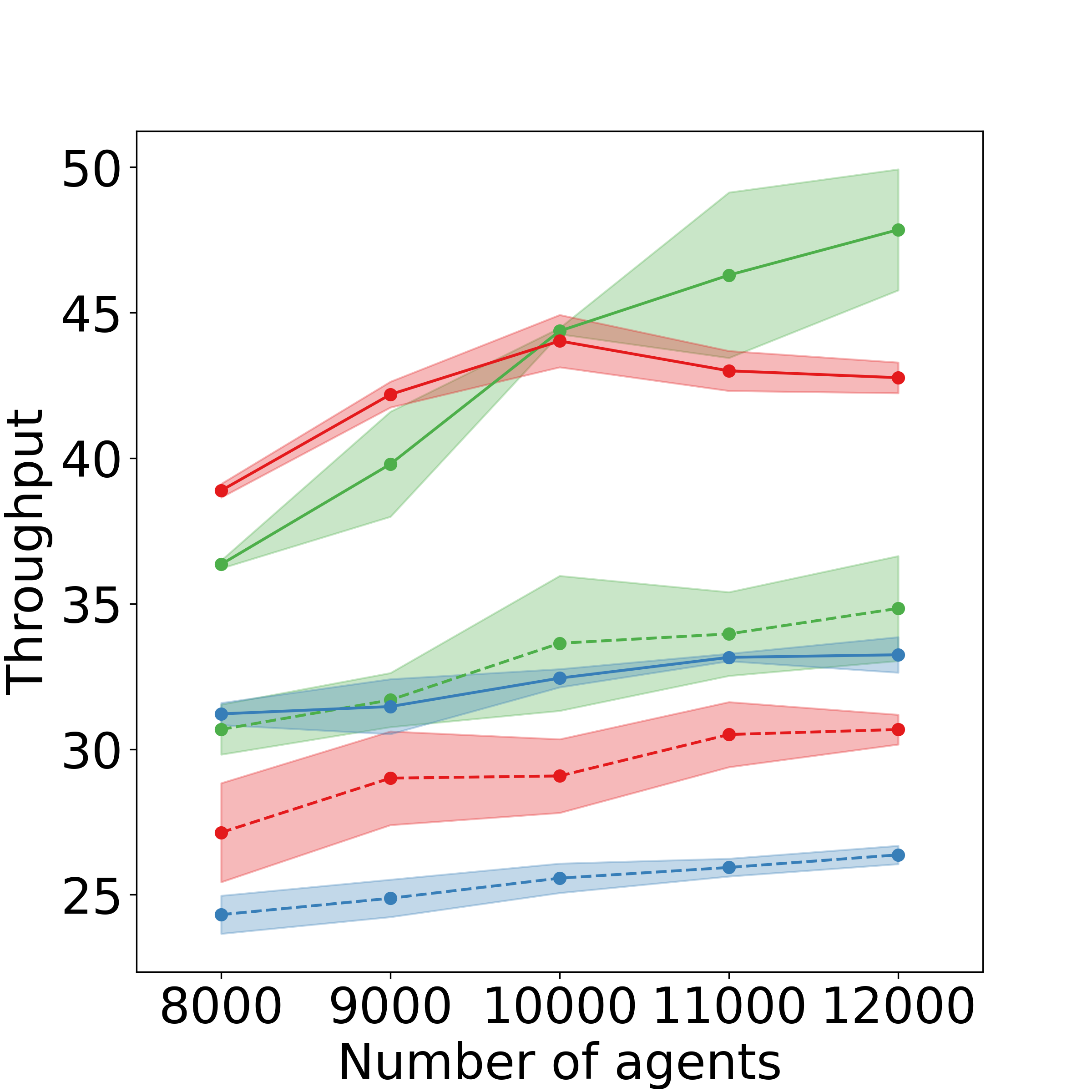}
      \caption{Random2}
      % \label{fig:random_corner}
    \end{subfigure}%
    
    \caption{Evaluation with different numbers of agents. We run each setting with $8$ different seeds. The shaded area represents the standard deviation of the throughput.}
    \label{fig: ablation_agent_numbers}
\end{figure}

In this section, we compare Learnable PIBT and PIBT with different global guidance and different numbers of agents in \Cref{fig: ablation_agent_numbers}. The conclusions are similar to the ones in \Cref{ss: experiments}. With the same global guidance, Learnable PIBT consistently outperforms PIBT, proving the effect of learning. Also, different global guidance excels in different scenarios. An interesting observation is that in map Random2, Backward Dijkstra (BD) performs better with $<10,000$ agents, and Dynamic Guidance (DG) performs better with $>10,000$ agents. The reason may be that with more agents, there is potentially more congestion, and DG addresses congestion better than BD.

\subsection{Evaluation on Learn-to-Follow Benchmark}
\label{ss: eval_learn_to_follow}
\begin{figure}[!b]
    \centering
    \begin{subfigure}[b]{0.45\textwidth}
      \centering
      \includegraphics[width=1\textwidth]{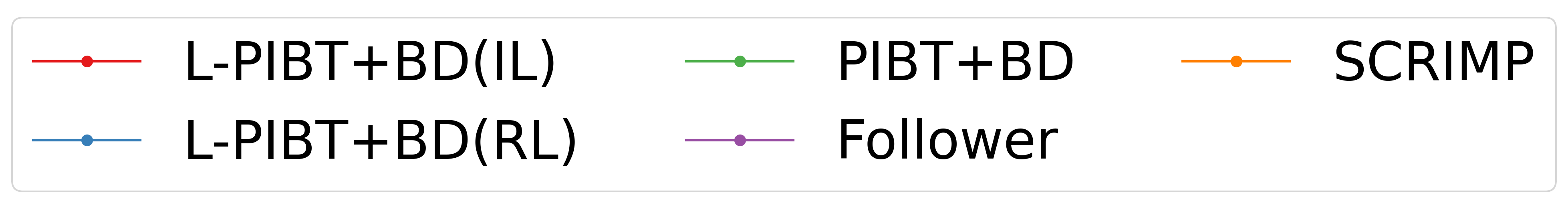}
      % \caption{Random 20*20}
      % \label{fig:random_corner}
    \end{subfigure}%

    \begin{subfigure}[b]{0.24\textwidth}
      \centering
      \includegraphics[width=1\textwidth]{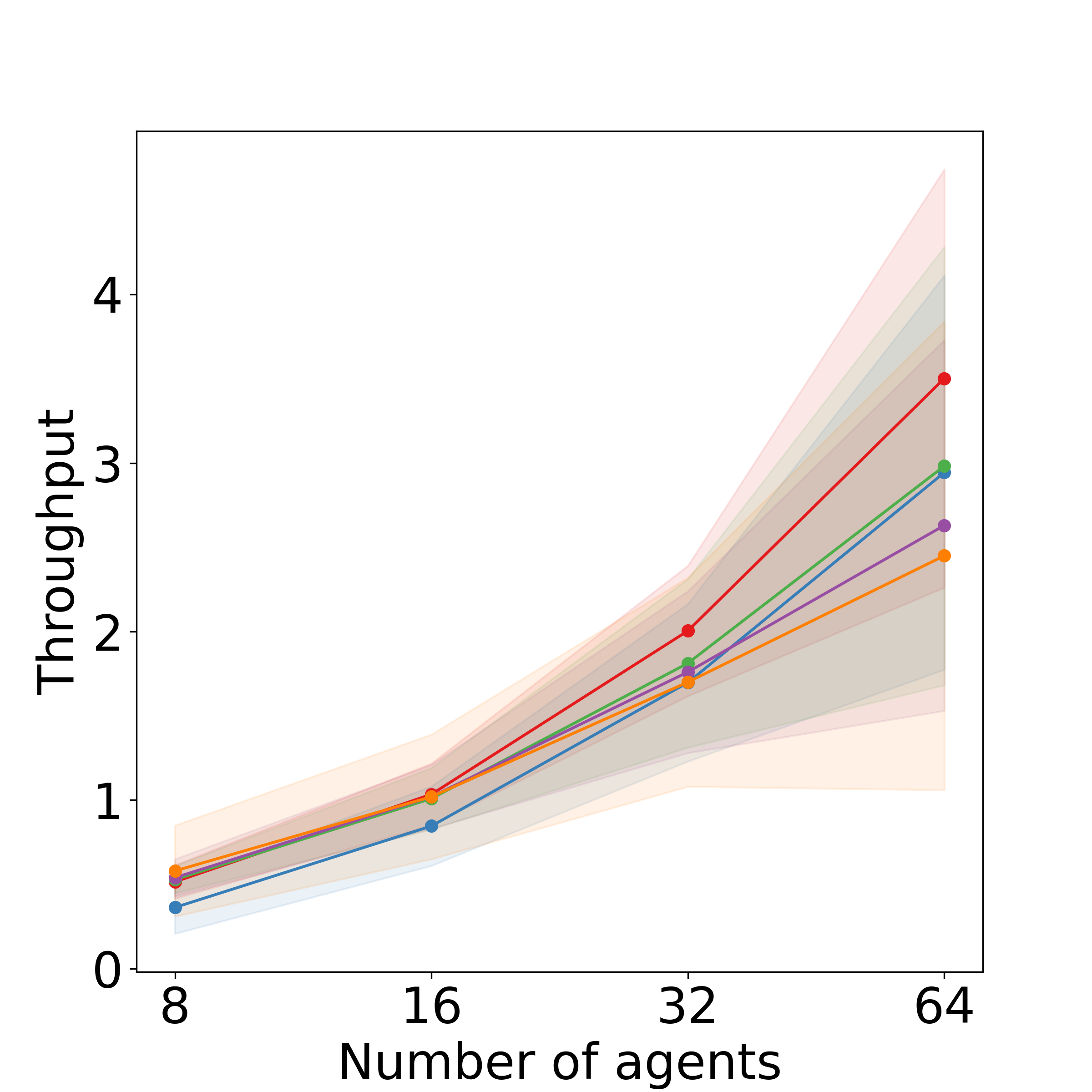}
      \caption{Random 20*20}
      % \label{fig:random_corner}
    \end{subfigure}%
    \hfill
    %\par\bigskip 
    \begin{subfigure}[b]{0.24\textwidth}
      \centering
      \includegraphics[width=1\textwidth]{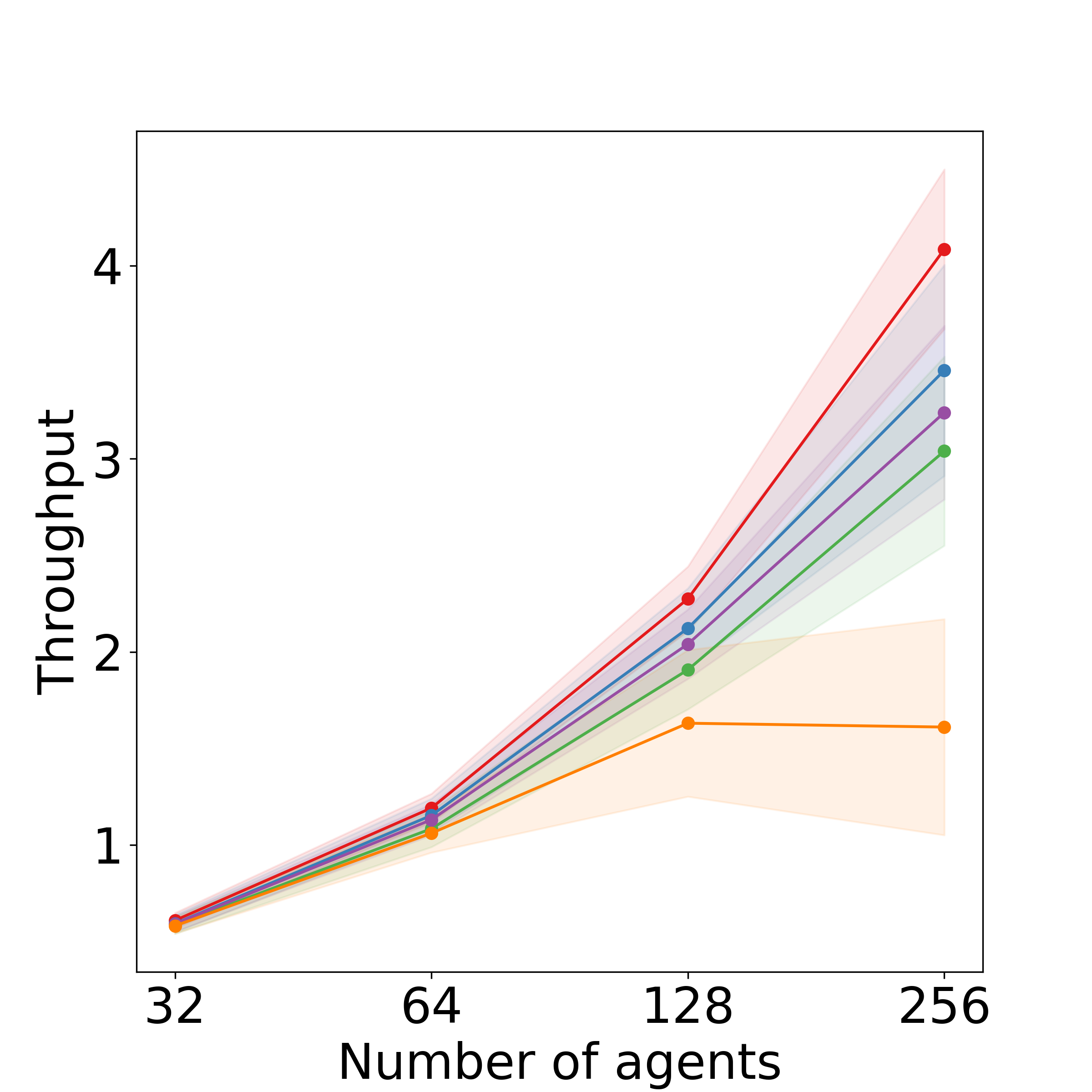}
      \caption{Mazes 65*65}
      % \label{fig:random_corner}
    \end{subfigure}%
    \hfill    %\par\bigskip 
    
    %\par\bigskip 
    \begin{subfigure}[b]{0.24\textwidth}
      \centering
      \includegraphics[width=1\textwidth]{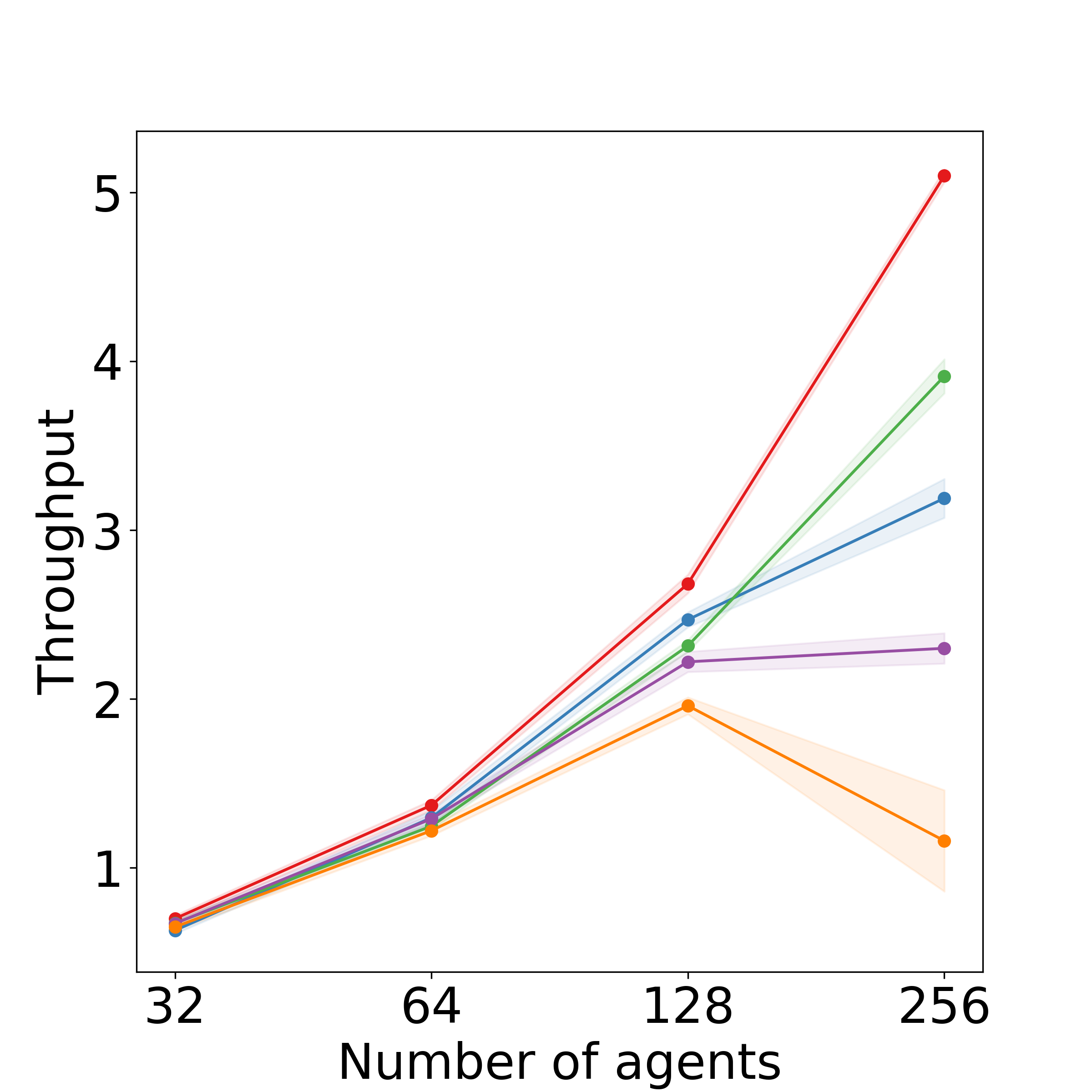}
      \caption{den520d 64*64}
      % \label{fig:random_corner}
    \end{subfigure}%
    \hfill
    \begin{subfigure}[b]{0.24\textwidth}
      \centering
      \includegraphics[width=1\textwidth]{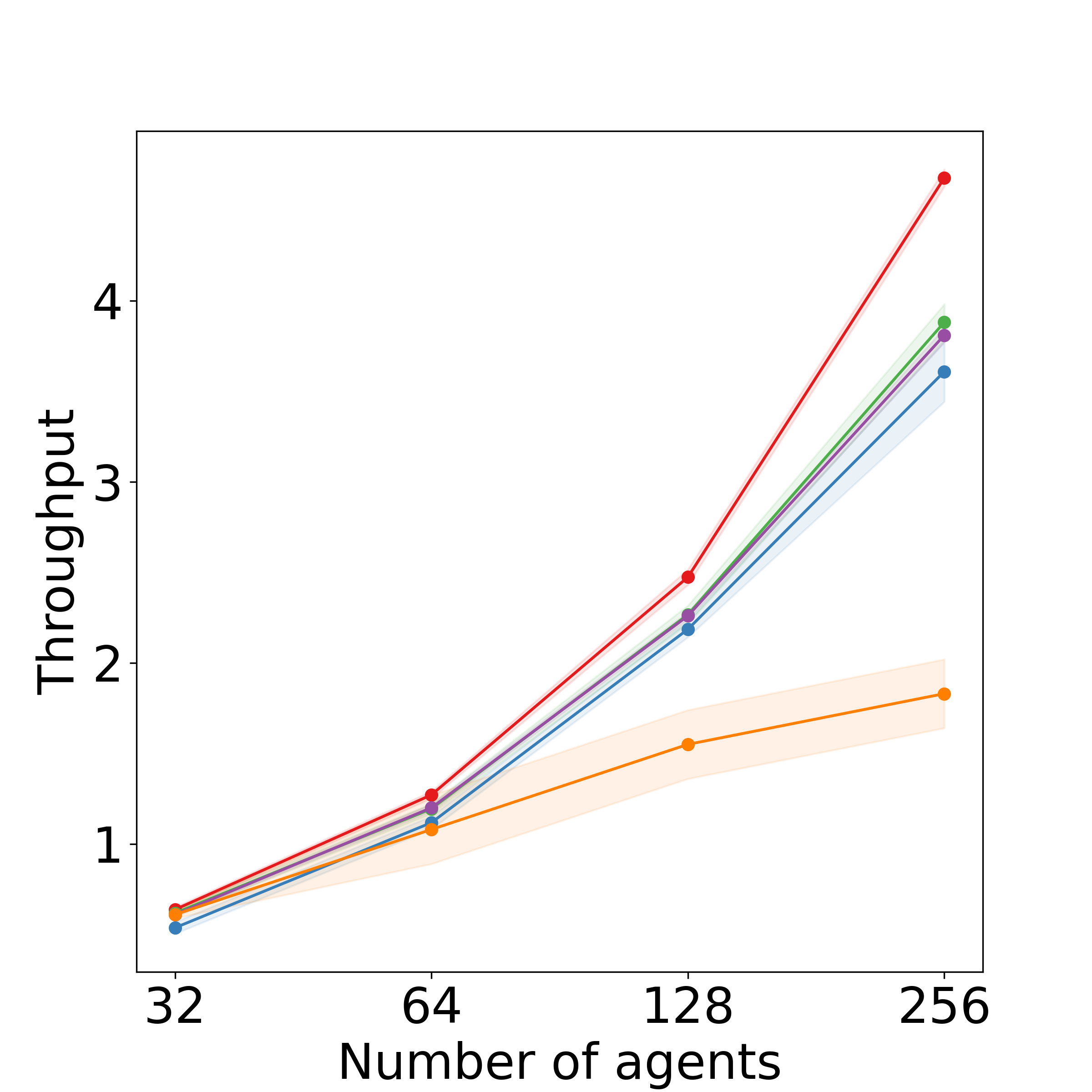}
      \caption{Paris\_1 64*64}
      % \label{fig:random_corner}
    \end{subfigure}%
    \hfill
    \caption{Evaluation on Learn-to-Follow Benchmark.}
    \label{fig: ablation_on_ltf_benchmark}
\end{figure}

This section compares different decentralized methods on the Learn-to-Follow Benchmark~\cite{skrynnik2024learn} to validate the superiority of our SILLM (Learnable PIBT) in \Cref{fig: ablation_on_ltf_benchmark}, as the experiment setting in the Learn-to-Follow paper is quite different from ours. Specifically, we compare Learnable PIBT trained with imitation learning and with reinforcement learning, Follower~\cite{skrynnik2024learn}, SCRIMP~\cite{wang2023scrimp}, and PIBT~\cite{okumura2022priority}. For a simple comparison, we only use Backward Dijkstra as global guidance. All the training settings are the same as the ones in the Follower paper~\cite{skrynnik2024learn}. Notably, Learnable PIBT and Follower are trained on $40$ Maze maps and then tested on $10$ different Maze maps and other types of maps. Our Learnable PIBT trained with imitation learning consistently performs the best across $4$ different types of maps. Notably, Follower actually only outperforms PIBT in Mazes maps but may fail to outperform PIBT in other maps, which means the generalization ability of Follower still needs improvement. In contrast, our Learnable PIBT is much more generalizable.

\subsection{Real-World Mini Example}

Since during the planning process, our algorithm assumes the position of all agents to be perfectly known at all times, we use ground truth positions for our virtual robots and use external localization (here, the Optitrack Motion Capture System) to obtain accurate position information for our real robots. However, the planned path may not be executed accurately due to disturbances and control inaccuracies. To eliminate these errors, we implement an Action Dependency Graph (ADG) \cite{berndt2023receding}. 

The video demo is available in the supplementary material. From our experiment with $10$ real agents, we observe that agents can reach their goals quickly without collisions, and errors are eliminated by the ADG, demonstrating the potential of using our method in the real world. In our experiment with $100$ virtual agents, we compare PIBT with Learnable PIBT. We can observe that Learnable PIBT outperforms PIBT with $50\%$ more throughput.

% % Appendixes should appear before the acknowledgment.

\subsection{Computation Resources}
Our models are trained on servers with 72 vCPU AMD EPYC 9754 128-Core Processor, 4 RTX 4090D (24GB), and 240 GB memory. Training on each map takes less than $12$ hours.

\subsection{Baseline Methods Implementation}

\subsubsection{WPPL}
In \Cref{ss: main results}, we compare our methods with the winning solution of League of Robot Runner Competition~\cite{chan2024league}, WPPL~\cite{Jiang2024Competition}. Our implementation is based on the public repo: https://github.com/DiligentPanda/MAPF-LRR2023. We remove the rotation action to align with the settings in other baselines. In addition, instead of limiting the planning time at each step to $1$ second, we limit the iterations of LNS refinement to $40,000$, roughly the total iterations used in our imitation learning. More iterations may improve the throughput, but the algorithm will run even slower as it has already been much slower than SILLM in \Cref{tab:main}.

\subsubsection{MAPPO}
In the ablation study (\Cref{ss: ablation study}), we show that Imitation Learning is better than simple MAPPO-based Reinforcement Learning~\cite{yu2022surprising}. Specifically, we use the following reward function.
\begin{align}
    r(v,v') = h(v)-h(v')-1
\end{align}
where $h$ is the heuristic function defined in \Cref{ss: heuristics}, $v$ and $v'$ are the current and next locations of an agent. Take Backward Dijkstra heuristics as an example. If $v'$ is $1$-step closer to the goal, the reward will be $0$. Otherwise, the reward will be a negative penalty. If no other agents act as obstacles, the agent should follow its shortest path given this reward function after learning. Reward design is crucial to the performance of RL and worth further study.

\subsubsection{Others}
For other baselines in \Cref{tab:main}, we directly use the implementation released by the corresponding papers. We re-trained models on our problem instances if they are learning-based methods.

\subsection{Detailed Comparison for Different Guidance}
% TODO
% Wait Heatmap (also gif)
% task curves
% Density Graph (also gif)
% We need to visualize why some guidance is better than others.

% We can define adaptiveness as the minimum score across all instances.

\begin{table*}[tbh]
\setlength{\tabcolsep}{3pt}
    \caption{The comparison of PIBT and L-PIBT with different guidance. The left part is the result of downscaled small instances. The right part is the result of the original large instances. We evaluate each instance with $8$ runs of different starts and goals. Each column records the mean throughput with the standard deviation in the parentheses. The Time (in seconds) and Score refer to the average single-step planning time and the average score defined in \Cref{ss: main results}. % ``-" indicates unavailable data due to the excessive planning time. We rank all the algorithms in descending order based on their average score on large-scale instances. The best throughput of each map is marked in bold. Notably, PIBT and TrafficFlow are exactly PIBT+BD and PIBT+DG in \Cref{fig: radar}. 
    }
    \label{tab:guidance}
    \centering
    \resizebox{0.95\textwidth}{!}{
    \begin{tabular}{c|cccccc|cc|cccccc|cc}
    % \toprule
    \toprule
        \multirow{2}{*}{Algorithm}  & \multicolumn{8}{c|}{Small maps with 600 agents} & \multicolumn{8}{c}{Large maps with 10,000 agents}\\
        %\cmidrule{2-14}
         & Sort & Ware & Pari & Berl & Ran1 & Ran2 & Time & Score & Sort & Ware & Pari & Berl & Ran1 & Ran2 & Time & Score \\
\midrule
PIBT \cite{okumura2022priority} & \makecell{7.79\\(0.36)} & \makecell{4.62\\(0.10)} & \makecell{5.59\\(0.15)} & \makecell{5.12\\(0.35)} & \makecell{10.84\\(0.08)} & \makecell{6.80\\(0.48)} & \makecell{0.004} & 0.60 & \makecell{32.44\\(0.10)} & \makecell{19.39\\(2.04)} & \makecell{15.43\\(0.34)} & \makecell{15.09\\(0.26)} & \makecell{46.51\\(0.08)} & \makecell{29.08\\(1.26)} & \makecell{0.014} & 0.61 \\
\midrule
L-PIBT & \makecell{14.76\\(0.38)} & \makecell{8.69\\(0.33)} & \makecell{7.76\\(0.17)} & \makecell{7.16\\(0.62)} & \makecell{\textbf{12.73}\\(0.11)} & \makecell{\textbf{10.76}\\(0.12)} & \makecell{0.007} & 0.89 & \makecell{42.91\\(0.07)} & \makecell{27.34\\(1.31)} & \makecell{18.84\\(0.20)} & \makecell{19.16\\(0.33)} & \makecell{\textbf{53.74}\\(0.10)} & \makecell{\textbf{44.02}\\(0.90)} & \makecell{0.024} & 0.80 \\
\midrule
PIBT+SG & \makecell{13.66\\(0.22)} & \makecell{9.91\\(0.24)} & \makecell{6.18\\(0.19)} & \makecell{4.50\\(0.74)} & \makecell{11.66\\(0.12)} & \makecell{7.46\\(0.37)} & \makecell{0.004} & 0.74 & \makecell{42.51\\(0.10)} & \makecell{39.34\\(0.11)} & \makecell{18.11\\(0.34)} & \makecell{17.62\\(0.26)} & \makecell{46.89\\(0.14)} & \makecell{25.57\\(0.51)} & \makecell{0.014} & 0.75 \\
\midrule
L-PIBT+SG & \makecell{\textbf{16.52}\\(0.09)} & \makecell{\textbf{14.28}\\(0.12)} & \makecell{\textbf{8.85}\\(0.11)} & \makecell{\textbf{7.19}\\(0.14)} & \makecell{12.34\\(0.11)} & \makecell{10.09\\(0.10)} & \makecell{0.007} & 0.98 & \makecell{\textbf{43.98}\\(0.07)} & \makecell{\textbf{42.24}\\(0.05)} & \makecell{22.54\\(0.17)} & \makecell{21.59\\(0.23)} & \makecell{52.49\\(0.10)} & \makecell{32.44\\(0.31)} & \makecell{0.023} & 0.85 \\
\midrule
\makecell{PIBT+DG\\(TrafficFlow \cite{chen2024traffic})} & \makecell{11.78\\(0.20)} & \makecell{9.10\\(0.42)} & \makecell{7.44\\(0.24)} & \makecell{5.70\\(0.43)} & \makecell{10.35\\(0.37)} & \makecell{8.48\\(0.10)} & \makecell{0.016} & 0.76 & \makecell{36.89\\(0.22)} & \makecell{27.78\\(0.88)} & \makecell{27.51\\(0.69)} & \makecell{27.03\\(0.54)} & \makecell{45.62\\(0.84)} & \makecell{33.64\\(2.32)} & \makecell{0.570} & 0.81 \\
\midrule
L-PIBT+DG & \makecell{14.41\\(0.16)} & \makecell{11.67\\(0.54)} & \makecell{8.34\\(0.07)} & \makecell{6.77\\(0.28)} & \makecell{11.14\\(0.13)} & \makecell{9.18\\(0.14)} & \makecell{0.029} & 0.88 & \makecell{39.38\\(0.09)} & \makecell{35.39\\(0.26)} & \makecell{\textbf{31.00}\\(0.89)} & \makecell{\textbf{30.52}\\(0.98)} & \makecell{50.48\\(0.14)} & \makecell{44.37\\(0.11)} & \makecell{0.721} & 0.94 \\
\midrule
    \end{tabular}
    }
\end{table*}

We report detailed throughput and average running time for comparing PIBT and L-PIBT with different guidance in \Cref{tab:guidance}. Notably, in contrast to the conclusion in the large map, Static Guidance works the best among the three heuristics in the small map for training. Overall, since all the heuristics are manually designed, the best heuristic is instance-dependent and should be evaluated empirically. Also, these heuristics only define the framework, but there could be a lot of hyperparameters that affect the performance. Automatic hyperparameter tuning can be helpful but doesn't necessarily remove the structural bias in the framework.

\end{document}